\documentclass[journal]{IEEEtran}
\usepackage{psfrag}
\usepackage{subfigure}
\usepackage{url}
\usepackage{stfloats}
\usepackage{cite}
\usepackage{amsmath}
\usepackage{amssymb}
\usepackage{amsfonts}
\usepackage{graphicx}
\usepackage{fancybox}
\usepackage{color}
\usepackage{algorithm}
\usepackage{algorithmic}
\usepackage{multirow}
\usepackage{booktabs}
\usepackage{threeparttable}
\usepackage{latexsym}
\usepackage{bm,array}
\usepackage{graphicx}
\usepackage{float}

\IEEEoverridecommandlockouts
\normalsize
\begin{document}

%%%%%%%%%%%%%%%%%%%%%%%%%%%%%%%%%%%%%%%%%%%%%%%%%%%%%%%%%%%%%%%%%%%%%%%%%%%%%
% Title and Authors
%%%%%%%%%%%%%%%%%%%%%%%%%%%%%%%%%%%%%%%%%%%%%%%%%%%%%%%%%%%%%%%%%%%%%%%%%%%%%
\title{5G Cellular User Equipment: \\From Theory to Practical Hardware Design}
\author{
Yiming~Huo,~\IEEEmembership{Student~Member,~IEEE},
Xiaodai~Dong,~\IEEEmembership{Senior Member,~IEEE},
and~Wei~Xu,~\IEEEmembership{Senior Member,~IEEE},\\
\thanks{Y. Huo, and X. Dong are with the Department of Electrical and Computer Engineering, University of Victoria, Victoria, BC V8P 5C2, Canada (ymhuo@uvic.ca, xdong@ece.uvic.ca).}
 \thanks{W. Xu is with the National Mobile Communications Research Laboratory (NCRL), Southeast University, Nanjing 210096, China (wxu@seu.edu.cn).}
}

\maketitle

\begin{abstract}
Research and development on the next generation wireless systems, namely 5G, has experienced explosive growth in recent years. In the physical layer (PHY), the massive multiple-input-multiple-output (MIMO) technique and the use of high GHz frequency bands are two promising trends for adoption. Millimeter-wave (mmWave) bands such as 28 GHz, 38 GHz, 64 GHz, and 71 GHz, which were previously considered not suitable for commercial cellular networks, will play an important role in 5G. Currently, most 5G research deals with the algorithms and implementations of modulation and coding schemes, new spatial signal processing technologies, new spectrum opportunities, channel modeling, 5G proof of concept (PoC) systems, and other system-level enabling technologies. In this paper, we first investigate the contemporary wireless user equipment (UE) hardware design, and unveil the critical 5G UE hardware design constraints on circuits and systems. On top of the said investigation and design trade-off analysis, a new, highly reconfigurable system architecture for 5G cellular user equipment, namely distributed phased arrays based MIMO (DPA-MIMO) is proposed. Finally, the link budget calculation and data throughput numerical results are presented for the evaluation of the proposed architecture.
\end{abstract}


\IEEEpeerreviewmaketitle
\newtheorem{mydef}{Definition}
\newtheorem{myLemma}{Lemma}
\newtheorem{theorem}{Theorem}
\newtheorem{Remark}{Remark}

\section{Introduction}

In the cellular world, tremendous efforts have been devoted to delivering higher quality of service (QoS) and quality of experience (QoE) since the very first cell phone call was made in 1973. In the 3rd Generation Partnership Project (3GPP) roadmap, several representative techniques have marked the milestones to further accelerate such trend, namely: device-to-device (D2D) communication for boosting geographic spectrum reusability \cite{Doppler:Device}; heterogeneous and small-cell network (HetSNet) targeting at deploying small cells in addition to macro cells at the same or different carrier frequencies \cite{Lopez-Perez:MassiveMIMO}; carrier aggregation (CA) for larger radio frequency (RF) bandwidths; new carrier type (NCT) for increasing spectral efficiency and spectrum flexibility, and reducing interference and power consumption \cite{Ericsson:Lean}; higher order modulation schemes and more layers of spatial multiplexing (SM) for higher spectral efficiency (SE). In 3GPP release 12, $8 \times8$ multiple-input-multiple-output (MIMO) for downlink, $4 \times4$ MIMO for uplink, 5 carrier components (CCs) and 256 quadrature amplitude modulation (QAM) are supported to satisfy the increasing wireless capacity needs. 

As part of the QoS requirement of 5G networks, the 5G peak downlink throughput (PDLT) is expected to achieve 10 Gbps in the dense urban environments \cite{Agiwal:Next}. Delivering such high speed data to end-users is an essential prerequisite for a satisfying QoE which is perceived subjectively. Furthermore, this high PDLT can be translated into a very high SE requirement of at least 100 bits/s/Hz based on the maximum 100 MHz bandwidth (BW) that a mobile network operator currently can support through enabling 5 CCs. If the RF bandwidth is fixed, such high SE requirement leads to using either higher order modulation scheme, more layers of SM, or both. The PDLT is given by
\begin{equation}\label{eq:pdlt}
\text{PDLT}\propto( B_{\text{RF}}\times{N_{\text{QAM}}}\times{N_{\text{CA}}}\times{N_{\text{MIMO}}} )  
\end{equation}
where $B_{\text{RF}}$ is the RF bandwidth for one single carrier, $N_{\text{QAM}}$ represents the modulation order, $N_{\text{CA}}$ is the number of aggregated carriers, and $N_{\text{MIMO}}$ stands for the number of MIMO spatial multiplexing layer. Therefore, the SE can be expressed as    
\begin{equation}\label{eq:se}
\text{SE}\propto( {N_{\text{QAM}}}\times{N_{\text{CA}}}\times{N_{\text{MIMO}}} ).  
\end{equation}
Nevertheless, from the implementation point of view, high modulation order and wide RF bandwidth unavoidably require power-hungry, complicated and high-performance RF and baseband circuits. On the other hand, high order of MIMO is confronted with the limitation of antennas' physical dimension, spacing, and radiation efficiency (RE). Based on emerging UE design techniques, a mobile phone handset can accommodate at most $4 \times4$ MIMO antennas with 256-QAM modulation \cite{Samsung}, which theoretically boosts PDLT up to approximately 1960 Mbps when using 5 CCs. On the other hand, such UE design already reaches the maximum spatial multiplexing gain due to the limited hardware area for embedding MIMO antennas of low GHz frequency bands. As a result, at the UE end, the highest achievable SE is around 20 bits/s/Hz. With the said SE, achieving 10 Gbps PDLT would require at least 500 MHz bandwidth, which is currently not possible in terms of the limited spectrum holdings of service providers.     
    
Meanwhile, Wi-Fi technologies have been advancing rapidly. The cutting-edge off-the-shelf IEEE 802.11ac compatible wireless products can support a $4 \times4$ multi-user MIMO (MU-MIMO) for downlink. By using unlicensed 60 GHz Millimeter-wave (mmWave) frequency bands, wireless gigabit alliance (WiGig) IEEE 802.11ad products deliver even higher data rate for short-range communications \cite{Samsung}. The yet to be released IEEE 802.11ax and 802.11ay standards that are deemed as the successors of 802.11ac and 802.11ad respectively, are expected to provide improved QoS such as better communication coverage and reduced latency. Besides WiFi, Bluetooth, near-field communication (NFC) and global navigation satellite system (GNSS) are also integrated in mobile handset terminals. These wireless technologies will compete with cellular for hardware resources and design budget on an already highly compact, multi-functional, multi-standard wireless handset terminal. For example, a huge challenge is to integrate antenna systems for different wireless technologies that occupy a very wide range of frequency (from 700 MHz to almost 6 GHz), and implementing high order MIMO can make it even more serious considering the limited dimension of a mobile handset device. 
    
In light of these challenging issues in sub-6 GHz, the high GHz frequency bands used to be considered unsuitable for commercial cellular networks are now attracting significant attention. On July 14, 2016, the Federal Communications Commission (FCC) voted to adopt a new Upper Microwave Flexible Use service in the licensed bands, namely 28 GHz (27.5-28.35 GHz), 37 GHz (37-38.6 GHz), 39 GHz (38.6-40 GHz), plus a new unlicensed band at 64-71 GHz \cite{FCC}. This initiative taken by the FCC helps mitigate the 5G UE design challenges. First, larger continuous RF bandwidth enables higher data rates. Second, using mmWave frequencies leads to a significant reduction of antenna dimension, and as a result, the form factor of the UE can be maintained while facilitating beamforming (BF) and spatial multiplexing.

In this paper, we present a novel distributed phased array MIMO (DPA-MIMO) architecture for 5G UE hardware design. The remainder of this paper is organized as follows, Section II investigates the contemporary wireless UE design and explains the design constraints. Section III presents the details of 5G UE hardware design challenges, and proposes a novel system architecture to address these challenges. Section IV conducts the performance evaluation of the proposed system based on the link budget calculation and comparison with state-of-the-art 5G works. Finally, Section V concludes this paper.    

\section{WIRELESS UE DESIGN OVERVIEW}
Contemporary wireless UE design becomes more challenging and complicated than ever. In any mainstream mobile phone, it does not only need to co-exist with several prevailing wireless technologies, but also integrate the camera(s), audio, battery, display, fingerprint scanner, vibrator, gyroscope, wireless charging, etc. In the near future, there will be a stronger need to enable or enhance various functions and technologies, such as virtual reality (VR), augmented reality (AR), internet of vehicles (IoV) \cite{Jiau:Multimedia}, and so on, all of which further increases the difficulty level of UE design. In this section, the major design methods and constraints are discussed from several aspects, namely battery, circuit and system, antenna and product design, and other system design trade-offs.

\subsection{Battery Design Constraints}
\begin{figure}
\centering
\includegraphics[scale = 0.95]{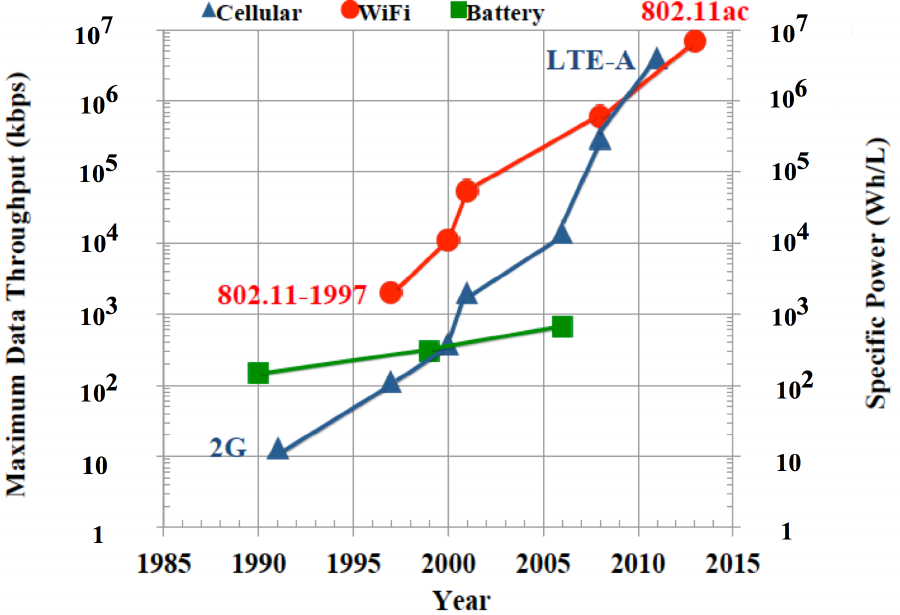}
\caption{Increment of the wireless capacity versus the battery performance improvement.}\label{fig:BATTERYFig}
\end{figure}

As plotted in Fig.~\ref{fig:BATTERYFig}, the data rates of both WiFi and cellular increase by around 10 folds for every five years. On the other hand, during the last 20 years, the battery technique for mobile devices has been through three major technical transitions, which starts with nickel-cadmium (NiCd) battery, then the nickel-metal hydride (NiMH) battery, and eventually the current mainstream lithium-ion (Li-ion) battery \cite{Reddy:Handbook}. From 1995 to 2014, the wireless capacity has increased by around 10,000 times \cite{Tsai:Cloud}, while only 4-5 folds increment of battery specific power have been achieved for the same period. Apparently, this mismatch becomes one of the current bottlenecks for mobile handset devices and affects the quality of user experience. Despite the recent battery research on new anode materials \cite{David:Silicon}, before the advent of significant breakthrough in battery performance and feasibility for mass production, higher energy efficiency of the UE wireless system will be critically relevant.
\subsection{Circuit and System Design}
The performance of a wireless system, from a hardware perspective, depends on the evolution of design arts in system-on-chip (SoC), printed-circuit board (PCB), mechanical design, and antenna design. The SoCs of high energy efficiency, small area, low cost and high yield, are always strongly desired. For the current SoC design, a widespread fact is that Moore's law slows down when the process dimension enters the deep-nanometer regime \cite{Sutardja:Slowing}. Consequently, the speed of energy efficiency improvement is moderated. Before any proven success with novel IC processes based on new materials, the contemporary silicon and III-IV compound based semiconductor processes, such as complementary metal-oxide-semiconductor (CMOS), CMOS silicon on insulator (SOI), fin field effect transistor (FinFET), silicon germanium (SiGe), gallium arsenide (GaAs), gallium nitride (GaN) and indium phosphide (InP), still play a dominant and critical role in the future 5G SoC designs.       

\begin{figure}
\centering
\includegraphics[scale = 0.95]{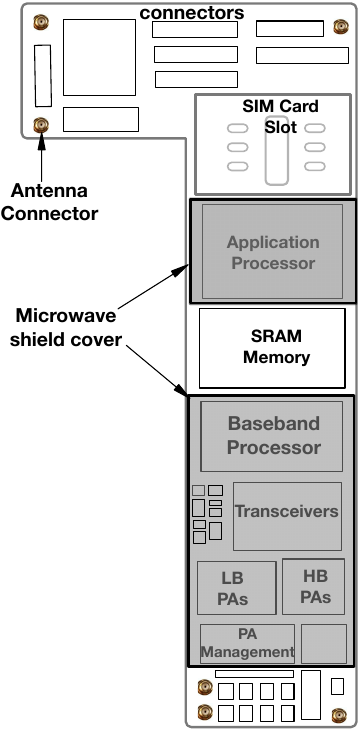}
\caption{An example of main logic board in contemporary smartphones.}\label{fig:MLBFig}
\end{figure}

Likewise, the multi-layer board design of a 5G mobile handset will become more compact and integrated to accommodate an increasing number of SoC chipsets for enabling various functions, standards, and technologies. On the main logic board (MLB) of a mobile handset as depicted in Fig.~\ref{fig:MLBFig}, there are cellular/WiFi RF transceivers, antenna switch modules, power amplifier (PA) modules, baseband (BB) modem, NFC, bluetooth, GNSS, application processor (AP), PA management unit, static random-access memory (SRAM), power management unit (PMU), etc. Nowadays, these highly customized chipsets are supplied by various vendors who design and fabricate them with different processes.

Similar to the trend in IC design, the footprint of PCB is continuously downsizing to smaller trace width and trace spacing. As a result, more chipsets can be embedded on one single main logic board, which results in less insertion loss (IL) and easier impedance matching. 
Therefore the RF front-end loss caused by IL and impedance mismatching are reduced, and the receiver (RX) sensitivity and transmitter (TX) power can be improved. On the other hand, signal integrity is an issue in a more complicated MLB design. For example, the clock signals and their harmonics, through complicated signal path and modulation, can end up at the receiver end in the form of spur. Therefore, a microwave shield cover is normally used on MLB to improve the electromagnetic compatibility/electromagnetic interference (EMC/EMI) performance. Another frequently seen issue is the degradation of sensitivity, or `desense', which is typically caused by TX output leaked into the RX path due to insufficient isolation between TX and RX ports. This issue is more serious in the case of carrier aggregation, for example, when the harmonic of a transmit signal falls in the receive band of a paired CA band \cite{Qorvo:CA}. As can be predicted, these issues will become more prevailing in a 5G terminal device. 

\subsection{Antenna and Product Design}
Antenna design is another matter of importance in wireless systems. Unlike any of its priors \cite{Rowell:Mobile} in the 2G/3G era, current mobile handset antennas are expected to support not only multi-bands and multi-standards in a wide range of frequency from 700 MHz to 6 GHz (with some uncovered gaps), but also enable certain degrees of diversity and SM. At the same time, there is requirement that high efficiency and low specific absorption rate (SAR) are both fulfilled after assembling antennas into the handset housing made of metallic casing.
       
Therefore, the co-design of antennas, metal casing, and handset housing is enormously challenging since the latter two factors could generate substantial effects on antenna performance \cite{Wong:IFA}. Narrow frame and metallic casing are still the unswerving trends currently and in the near future, because they enable better protection, portability, heat dissipation and aesthetic appearance. The slim form factor improves the user experience, and can be seen in several recent mainstream smartphones as shown in Table~\ref{tab:dimension}.  
   
\begin{table}[h]
\caption{Dimension information of smartphones.} \label{tab:dimension}
\newcommand{\tabincell}[2]{\begin{tabular}{@{}#1@{}}#2\end{tabular}}
 \centering
 \begin{tabular}{|c|c|c|c|c|}\hline
        \tabincell{c}{ Model }  & \tabincell{c}{ Dimension\\ ( ${H}\times{W}\times{D}$, mm) }  & \tabincell{c}{Display\\ Size (inch)} & \tabincell{c}{ Weight\\ (g) } \\  \hline
        \tabincell{c}{Samsung\\S7 edge} & \tabincell{c}{ $150.9\times72.6\times7.7$} & 5.5 & 157    \\
        \hline
        \tabincell{c}{Apple\\iPhone 7 plus} & \tabincell{c}{ $158.2\times77.9\times7.3$} & 5.5 & 188    \\
        \hline
        \tabincell{c}{Huawei\\Mate 9} & \tabincell{c}{ $156.9\times78.9\times7.9$} & 5.9 & 190    \\
        \hline     
        \tabincell{c}{Google\\Pixel XL} & \tabincell{c}{ $154.7\times75.7\times7.3$} & 5.5 & 168     \\\hline
    \end{tabular}
\end{table}

The antenna dimension is proportional to the effective wavelength, and this relation can be approximated as 
\begin{equation}\label{eq:WAVE}
{{\lambda }_{e}}\text{=}\frac{{{c}_{0}}}{f\sqrt{{{\varepsilon}_{e}}}}\
\end{equation}
where $c_{\text{0}}$ is the speed of light in vacuum, $f$ is the frequency, and ${\varepsilon}_{e}$ is the effective dielectric constant that makes the effective wavelength shorter. Furthermore, the effective dielectric constant can be derived using the following equations \cite{Balanis:Antenna}
\begin{equation}\label{eq:DiEff1}
\begin{aligned}
\ { {\varepsilon}_{e}}\text{=}\frac{{\varepsilon}_{r}+1}{2}+\frac{{\varepsilon}_{r}-1}{2}\Bigg\{\frac{1}{\sqrt{1+12\left(\frac{H}{W}\right)}} \\ 
 \hspace{1.5cm} +0.04{{\left( 1-\left( \frac{W}{H} \right) \right)}^2} \Bigg\}, \\
 \text{  subject to } W/H <1  
\end{aligned}
\end{equation}

\begin{equation}\label{eq:DiEff2}
\begin{aligned}
\ { {\varepsilon}_{e}}\text{=}\frac{{\varepsilon}_{r}+1}{2}+\frac{{\varepsilon}_{r}-1}{2\sqrt{1+12\left(\frac{H}{W}\right)}} \\  
\text{subject to } W/H \geq1  
\end{aligned}
\end{equation}
where ${\varepsilon}_{\text{r}}$ stands for the relative dielectric constant, $W$ is the width of antenna, and $H$ is the thickness of the antenna substrate. Therefore, the antenna dimension is mainly determined by the frequency and substrate material. Although higher dielectric constant reduces the antenna dimension, it degrades the antenna performance as more radiation energy will be confined inside the substrate instead of being radiated. 

In the 5G era, the handset antenna design faces more challenges in order to cover the legacy 3GPP standards and new 5G standards which regulate the use of higher GHz frequency bands. From this point of view, implementing large scale MIMO at low GHz frequencies becomes very difficult as it normally requires a minimum spacing to guarantee good isolation. As for the mmWave antenna design, more antenna elements can be accommodated thanks to downsizing, but the metal casing can deteriorate the antenna performance.
\begin{table*}[h]
\caption{Calculation and comparison of path loss.} \label{tab:Pathloss}
\newcommand{\tabincell}[2]{\begin{tabular}{@{}#1@{}}#2\end{tabular}}
 \centering
 \begin{threeparttable}
 \begin{tabular}{|c|c|c|c|c|c|c|}\hline
        \tabincell{c}{} & \multicolumn{6}{c|}{\textbf{Path loss of popular deployment scenarios (dB)}}      \\  \hline
        \tabincell{c}{\textbf{Frequency/distance}}  & \tabincell{c}{ UMa-LOS }  & \tabincell{c}{UMa-NLOS} & \tabincell{c}{UMi-Street\\Canyon-LOS} &\tabincell{c}{UMi-Street\\Canyon-NLOS} &\tabincell{c}{UMi-Street\\Open-LOS} & \tabincell{c}{UMi-Street\\Open-NLOS} \\  \hline
        \tabincell{c}{LTE Band 41\\2.6 GHz, d=100m} & \tabincell{c}{84.8} & \tabincell{c}{107.5} & \tabincell{c}{83.4} & \tabincell{c}{112.7}  & \tabincell{c}{81.9} &\tabincell{c}{105.6}  \\ 
        \hline
        \tabincell{c}{28 GHz, d=100m} & \tabincell{c}{105.5} & \tabincell{c}{128.2} & \tabincell{c}{104.1} & \tabincell{c}{133.4}  & \tabincell{c}{102.6} &\tabincell{c}{126.3}  \\ 
        \hline
        \tabincell{c}{39 GHz, d=100m} & \tabincell{c}{108.4} & \tabincell{c}{131.1} & \tabincell{c}{107} & \tabincell{c}{136.3}  & \tabincell{c}{105.5} &\tabincell{c}{129.2}     \\
        \hline     
        \tabincell{c}{39 GHz, d=100m,\\rain and oxygen loss\tnote{1}} & \tabincell{c}{109.4} & \tabincell{c}{132.1} & \tabincell{c}{108} & \tabincell{c}{137.3}  & \tabincell{c}{106.5} &\tabincell{c}{139.2}     \\
        \hline 
        \tabincell{c}{LTE Band 41\\2.6 GHz, d=1 km} & \tabincell{c}{104.9} & \tabincell{c}{137.5} & \tabincell{c}{103.2} & \tabincell{c}{144.6}  & \tabincell{c}{100.4} &\tabincell{c}{134.5}     \\
        \hline 
        \tabincell{c}{28 GHz, d=1 km} & \tabincell{c}{125.5} & \tabincell{c}{158.2} & \tabincell{c}{123.9} & \tabincell{c}{165.3}  & \tabincell{c}{121.1} &\tabincell{c}{155.2}  \\ 
        \hline
        \tabincell{c}{39 GHz, d=1 km} & \tabincell{c}{128.4} & \tabincell{c}{161.1} & \tabincell{c}{126.8} & \tabincell{c}{168.2}  & \tabincell{c}{124} &\tabincell{c}{158.1}     \\
        \hline     
        \tabincell{c}{39 GHz, d=1 km,\\rain and oxygen loss\tnote{1}} & \tabincell{c}{136.5} & \tabincell{c}{169.2} & \tabincell{c}{134.9} & \tabincell{c}{176.3}  & \tabincell{c}{132.1} &\tabincell{c}{166.2}        \\\hline    
    \end{tabular}
    \begin{tablenotes}
        \footnotesize
        \item[1] Heavy rain of 25mm/h model is used
      \end{tablenotes}
    \end{threeparttable}
\end{table*}

\subsection{System Design Trade-offs}
\begin{figure}
\centering
\includegraphics[scale = 1.0]{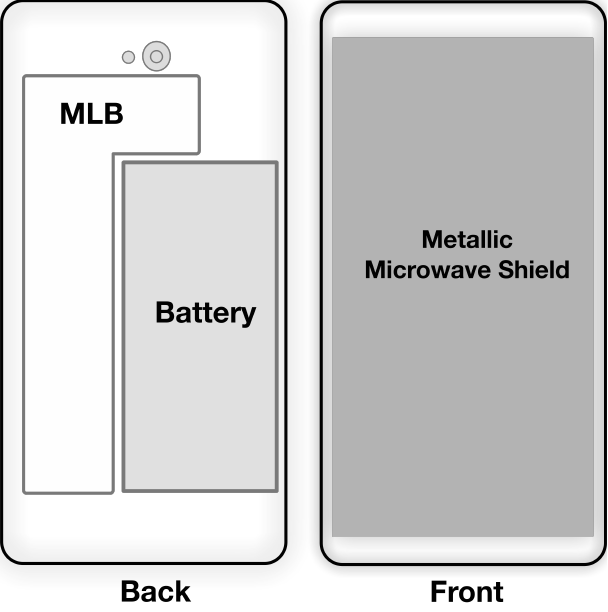}
\caption{Disassembly of a smartphone to front and back parts.}\label{fig:DISSAMBLYFig}
\end{figure}

\begin{figure}
\centering
\includegraphics[scale = 1.0]{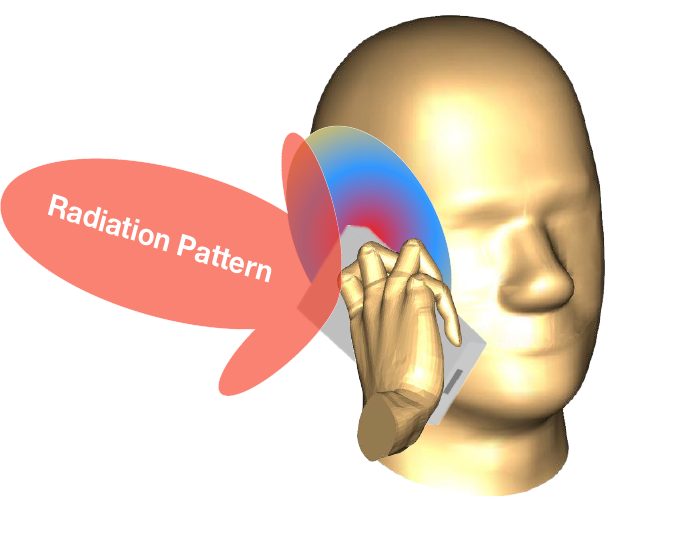}
\caption{Talk mode using a specific anthropomorphic mannequin (SAM) head phantom.}\label{fig:TALKFig}
\end{figure}

In addition to the aforementioned three major design considerations, there are also high-level design constraints between the wireless subsystem and other UE components. Besides the power budget and hardware area allocation, one more critical technical challenge originates from the interference among different components. For example, the display screen can cause the RF sensitivity degradation. Therefore, a sheet of metallic microwave (MW) shield is normally put between the display unit and hardware part to enhance the isolation as shown in Fig.~\ref{fig:DISSAMBLYFig} which briefly depicts a cell phone opened from middle. Moreover, this MW shield can minimize the SAR in the common use cases when the screen side is held close to the head of a smartphone user, as illustrated in Fig.~\ref{fig:TALKFig}.

In other words, antennas radiate minimal signal through the screen, and therefore it can only propagate the signal in the direction away from the human head. Nevertheless, the shield increases the thickness of the handset and degrades the form factor. The placements of camera, speaker, finger scanner, battery, MLB, also require careful consideration as they can change the electro-magnetic (EM) field and lead to undesired effects. To summarize here, contemporary wireless UEs need to provide high quality of user experience determined and contributed by comprehensive factors which not only lie in the wireless system design, but also mechanical design, product design, operating system design, etc. Consequently, many design trade-offs must be considered for a high-performance 5G UE. By taking the cellular standard as an example, the figure-of-merit (FOM) of a cellular UE can be formulated as
\begin{equation}\label{eq:FOM}
\begin{aligned}
&\ \text{FOM}_{\text{Cellular,UE}}= \\
&\ \frac{\sum\nolimits_{n=1}^{n,max}\frac{ \text{PDLT}_\text{Non-CA}^\text{Band,n} }{ {B}_\text{eff,n}{{P}_\text{n}} }+\sum\nolimits_{m=2}^{m,max}\frac{ \text{PDLT}_\text{CC,m}^\text{Bands} }{ {B}_\text{eff,m}{{P}_\text{m}} }}
{ {V}_\text{UE}\cdot{M}_\text{UE}}  
\end{aligned}
\end{equation}
where $\text{PDLT}_\text{non-CA}^\text{Band,n}$ is the $\text{PDLT}$ of the 3GPP band $n$ when carrier aggregation is not enabled (non-CA). ${B}_\text{eff,n}$ and ${P}_\text{n}$ stand for the effective bandwidth and power consumption respectively, when the wireless UE works in the non-CA mode. Moreover, the effective bandwidth is the bandwidth of the used band which has excluded the guard band. Accordingly, $\text{PDLT}_\text{CC,m}^\text{Bands}$ represents the $\text{PDLT}$ of the carrier aggregation of $m$ CCs, and the superscript $m,max$ is the maximum number of CCs, defined to be up to 5 in 3GPP Release 13. Thus, the first and second terms of the numerator add up the energy-spectral efficiency of both non-CA and CA cases for all cellular bands and CA combinations supported by the wireless UE, then we divide the result by the volume ${V}_\text{UE}$ and weight ${M}_\text{UE}$ of the wireless UE. The denominator reflects the `score' of electrical-mechanical co-design and the portability of the wireless UE. Therefore the unit of $\text{FOM}_{\text{Cellular,UE}}$ is bit/Hz/Joule/$\text{mm}^3$/$\text{gram}$. It is obvious that more bands and CCs, higher SE, smaller volume and weight, can result in a higher FOM of wireless UE, which means a better comprehensive design. It is worth mentioning that the material cost of the UE is not taken into consideration in the equation for well-defined comparison.

\section{5G CELLULAR UE BASED ON A NOVEL SYSTEM ARCHITECTURE}
The foremost challenge of using high GHz frequency bands comes from the propagation loss that is significantly higher and more complicated than sub-6 GHz frequency bands. Based on the most recently published 5G channel model \cite{NYU:5G}, \cite{Haneda:5G}, plus atmospheric absorption and rain attenuation models in \cite{Rappaport:Millimeter}, \cite{Zhao:Rain}, the path loss comparison for different propagation scenarios are given in Table~\ref{tab:Pathloss} for three frequency bands, namely 2.6 GHz, 28 GHz, and 39 GHz. 
\subsection{Channel Model Analysis}
As presented in Table~\ref{tab:Pathloss}, the path loss of non-line-of-sight (NLOS) is much larger than that of line-of-sight (LOS), and LOS of the urban macro (UMa) scenario has similar path loss to the LOS urban microcell (UMi) scenario. However, the path loss in UMi Street Canyon NLOS is much more severe than UMa NLOS or UMi Street Open.  In addition, for all scenarios, the path loss of 28 and 39 GHz are at least 20 dB larger than LTE band 41. In order to combat such large path loss, the FCC regulation allows base station (BS) to transmit at 75 dBm per 100 MHz \cite{FCC}. Moreover, the power loss caused by oxygen absorption and rain attenuation are comparatively small. 

Apart from the path loss and atmospheric or rain attenuation loss, the building penetration loss depends on different materials. Particularly for the concrete wall, the penetration loss significantly increases with frequency \cite{Haneda:5G}, and it can be as high as 117 dB for 28 GHz. In light of these challenges, beamforming is mandatory at both the BS and UE end.

\subsection{Novel Distributed Phased Array Based MIMO Architecture}
Implementing mmWave beamforming at the UE end is more difficult than at the BS end since it is largely constrained by high energy efficiency requirement, limitations in battery life and hardware dimension which are key FOM contributors. The conventional concept of BF is a method to increase the signal-to-noise-ratio (SNR) and reduce channel interference of a single data stream, but does not provide spatial multiplexing gain by delivering multiple streams. 

For the antenna array design of a BF module, the spacing between antenna elements is critically important because it is directly relevant to the grating lobe when the array operates the beam steering. A grating lobe, or side lobe is undesired in the antenna array because it degrades the gain and radiation efficiency of an array \cite{TED:MMWAVE_BOOK}. Furthermore, for a given maximum amount of beam steering $\theta_\text{max}$, the spacing $d$ between two neighboring antenna elements has to be maintained as \cite{TED:MMWAVE_BOOK}
\begin{equation}\label{eq:spacing}
  \frac{d}{\lambda_\text{0}} \leq \frac{1}{1+cos\theta_\text{max}}, 
\end{equation}
where $\lambda_\text{0}$ is the free space wavelength. Furthermore, a maximum spacing of $\lambda_\text{0}$/2 is normally used to avoid grating lobes.

On the other hand, in a MIMO system, enough separation between two antenna elements needs to be maintained so that a good spatial multiplexing gain can be obtained. The adjacent element spacing depends on the specific antenna array design and use case. In order to avoid significant capacity degradation, adjacent element spacing should be larger than 1.5$\lambda_\text{0}$ for the uniform square array (USA)\cite{Rusek:MIMO}.  	     

Therefore, the design principles for antenna arrays used for beamforming and MIMO antennas used for spatial multiplexing are quite different. It is not feasible and practical to make an M$\times$N-element antenna array which is designed and optimized for beamforming to be perfectly used as M$\times$N MIMO antennas and vice versa.   

Consequently, at the UE end, a new system architecture needs to be proposed to enable the functionality including both beamforming and spatial multiplexing, for different application scenarios.

\begin{figure}
\centering
\includegraphics[scale = 0.52]{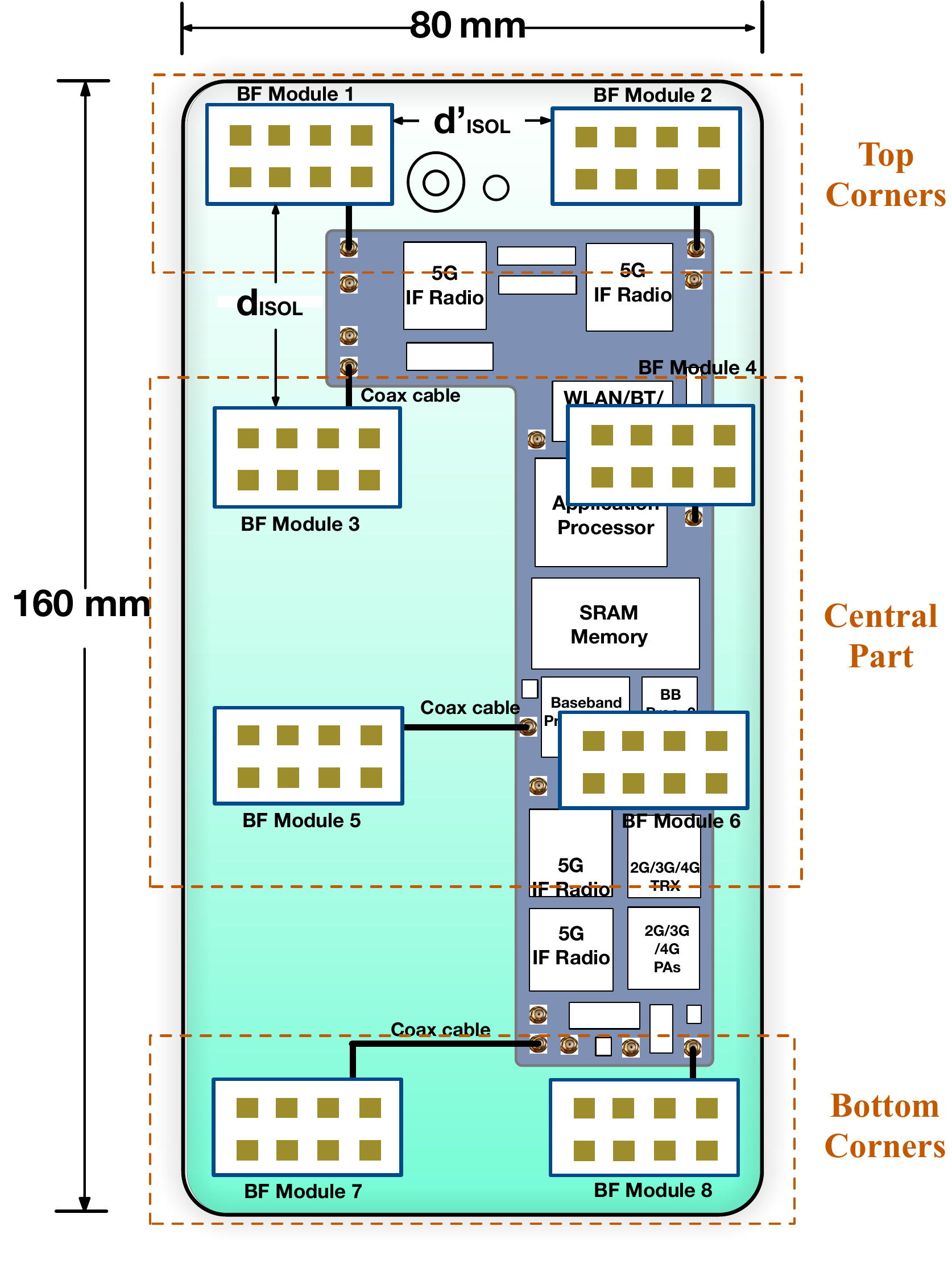}
\caption{Proposed DPA-MIMO architecture in mobile phone handset from back side transparent view.}\label{fig:5GUEFig}
\end{figure}  

\begin{figure*}[!t]
\centering
\includegraphics[scale = 0.85]{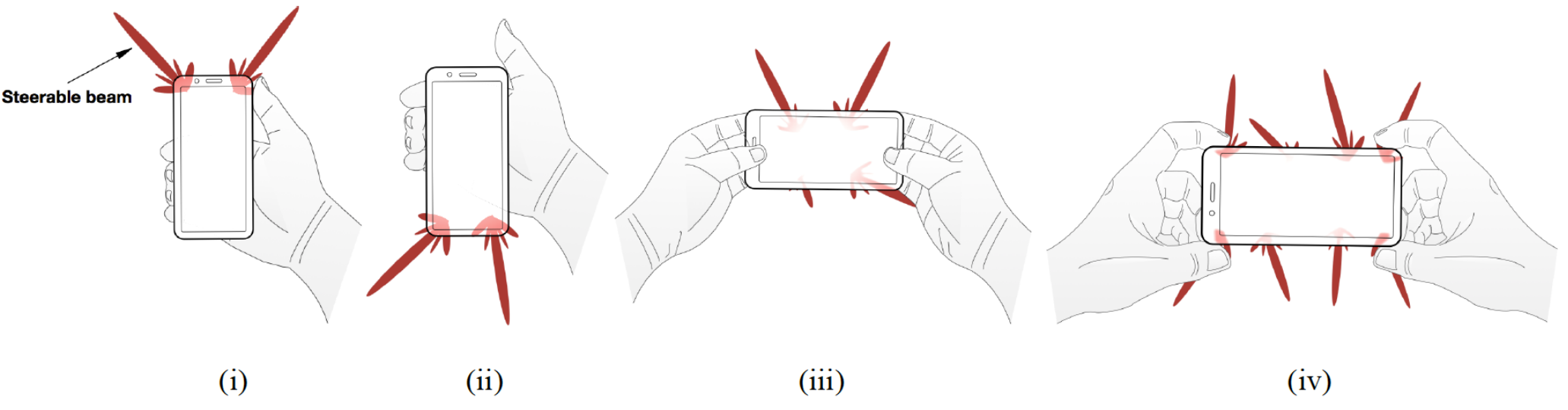}
\caption{Four popular positions of holding mobile phone handsets.}\label{fig:HoldFig}
\end{figure*}

Assuming the appearance of 5G mobile phone is similar to the emerging 4G ones, for example, the anticipated volume is around $160\times80\times8$ mm, and the SIM card slot is removed due to the use of embedded subscriber identity module (eSIM) or virtual SIM that can be integrated into a chipset. To be more specific, eSIM is designed and implemented by the UE manufacturer, and it removes the need of physical SIM card while easing the process of switching service provider \cite{Apple:eSIM}. It makes the UE device neat and user-friendly by getting rid of the physical SIM card tray, which is critical to not only the 5G UE but also the internet of things (IoT) devices such as wearable devices. 

A proposed 5G prototyping hardware design is illustrated in Fig.~\ref{fig:5GUEFig} where eight identical 8-element phased array based BF modules are distributed and placed in the back housing of a mobile handset. This new architecture is referred to as the distributed phased arrays based MIMO (DPA-MIMO) architecture, several advantages can be observed as follows.

\begin{figure*}
\centering
\subfigure [] {\includegraphics[scale=0.44]{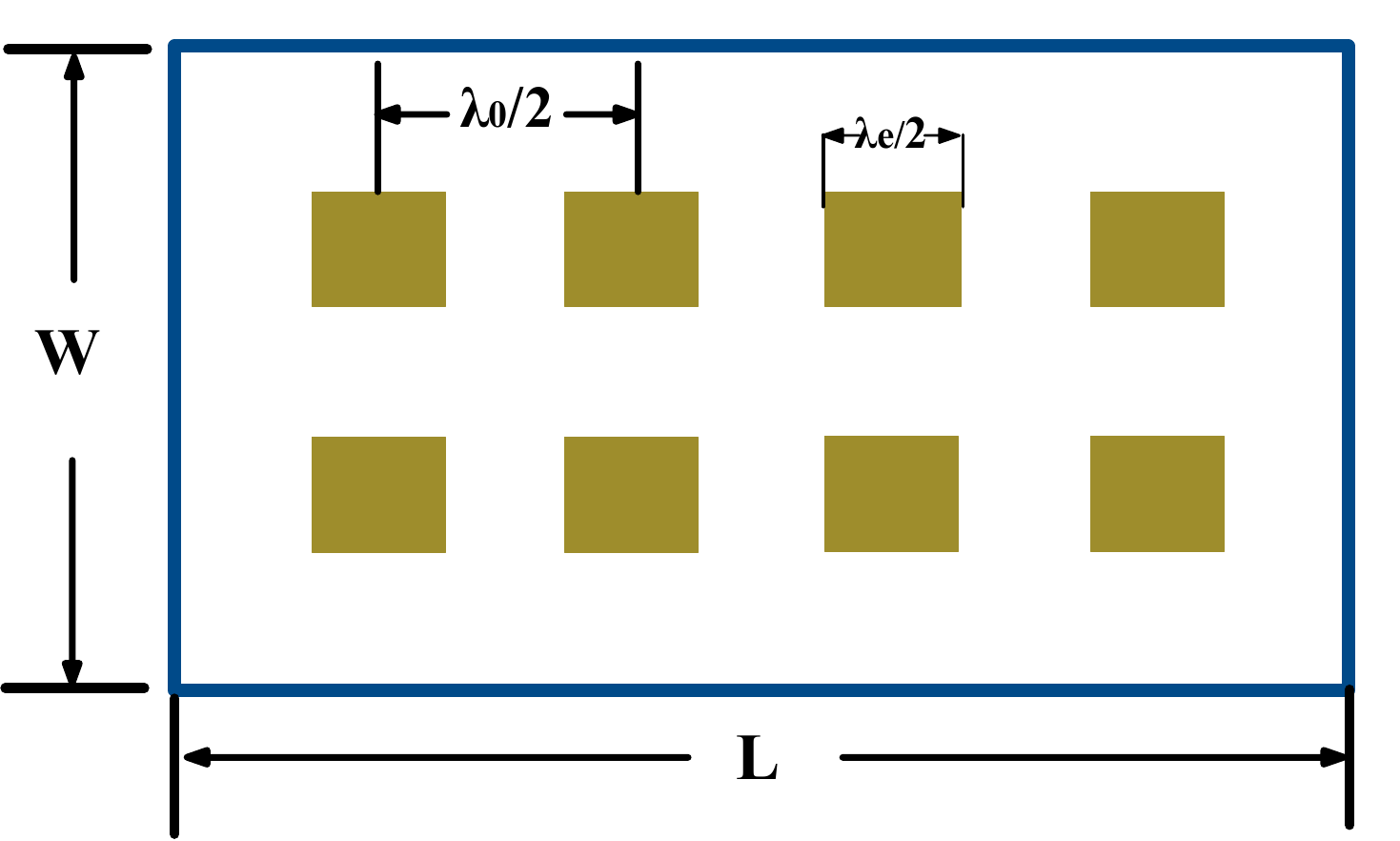}} \hspace*{ 1.4cm}\label{fig:BF_Fig}
\subfigure [] {\includegraphics[scale=0.36]{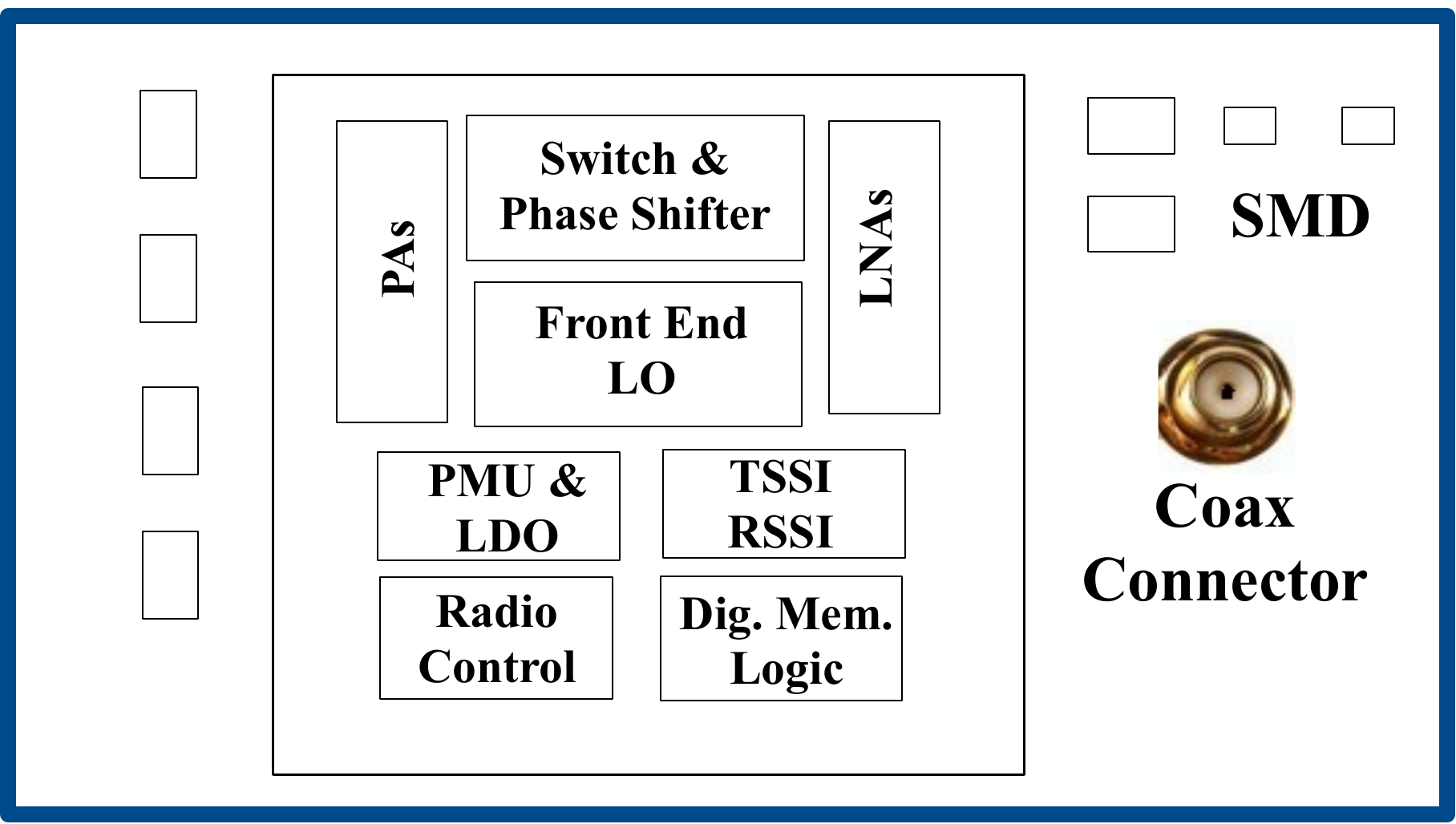}} \label{fig:BF_ChipFig}
\subfigure [] {\includegraphics[scale=0.75]{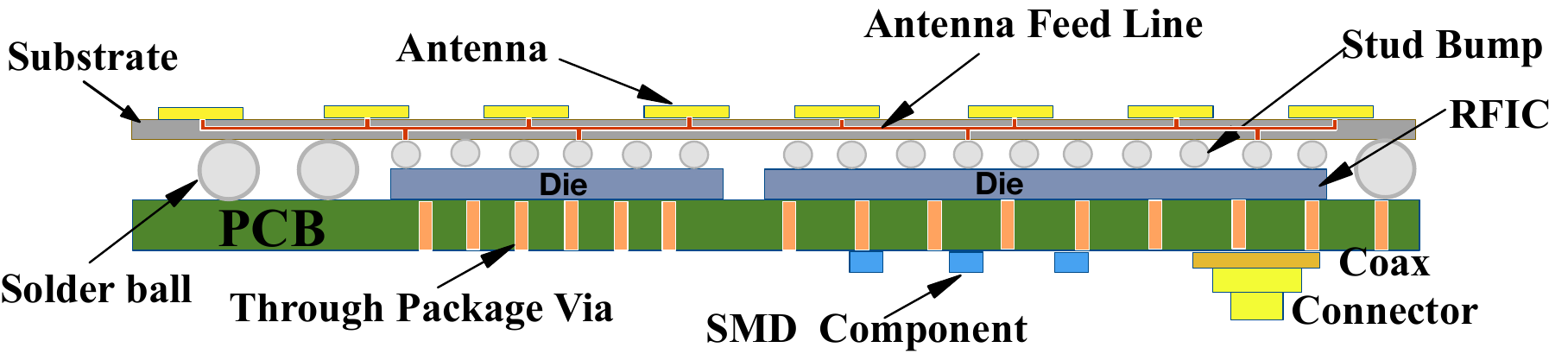}}\label{fig:BF_CrossFig}
\caption{(a) Top-down view of the BF module. (b) Chipsets layer of the BF module. (c) Layout and cross-section view of the BF module stack-up.}\label{fig:BF_Fig}
\end{figure*}

\begin{enumerate}
 \item Each BF module, embedding one RF transceiver chain, realizes an active phased array of $N_\text{ANT}$ (=8 in this example) antenna elements. Now we can use an example to estimate the effective isotropic radiated power (EIRP). Assume that a PA has an output power of $P_\text{PA}$, and the output power is split into $N_\text{ANT}$ equal parts and fed into a $N_\text{ANT}$-element phased array. At the phased array output, we can obtain an EIRP which is $10\text{log}_\text{10}(N_{\text{ANT}})$ (=9) dB higher than $P_\text{PA}$. Furthermore, if we remove that PA and place one PA in the front-end path of each antenna element, and the output power of all PAs is maintained at $P_\text{PA}$ (not scaled down with the increased number of front-ends \cite{Boers:60 GHz}), the EIRP can be boosted to be $20\text{log}_\text{10}(N_{\text{ANT}})$ (=18) dB higher than $P_\text{PA}$. Therefore, in this scenario, $N_\text{PA}$ (=$N_\text{ANT}$) PAs contribute the extra $10\text{log}_\text{10}(N_{\text{ANT}})$ (=9) dB gain on top of the first EIRP, but at the cost of higher power consumption.\\

 \item A total number of $N_\text{BF}$ (=8 in this example) BF modules, with enough spacing and isolation, can also cooperate as 8 MIMO antennas to process a maximum number of 8 streams. Thus, the spatial multiplexing gain can be obtained to further increase the link throughput by multiple times.\\

 \item The DPA-MIMO topology provides a solution to human body blockage which could lead to severe attenuation at mmWave frequencies \cite{Zhao:User}. For example, the attenuation can be as high as 30 to 40 dB for the 73 GHz band as given in \cite{MacCartney:Millimeter}. According to the study of mobile phone user habits \cite{Lau:Antenna}, the mobile handset is usually held in several popular positions as depicted in Fig.~\ref{fig:HoldFig}. In position (i) and (ii), thanks to the DPA-MIMO architecture, BF modules 1 and 2 can still work normally, either independently or cooperatively in a $2\times2$ MIMO SM mode; in case (iii), BF modules 3-6 can work either independently or cooperatively in a $4\times4$ MIMO SM mode. Finally, in position (iv), all BF modules can work simultaneously and support a $8\times8$ MIMO SM mode.\\ 
 
 \item Based on the design methodologies and consideration discussed above, placing BF modules at top two corners, bottom two corners, and the central part of the mobile device is mandatory in order to overcome the human body (hand) blockage issue. Therefore, $N_\text{BF}$ is flexible but has to be more than 5 as long as it satisfies the minimum isolation spacing. It is worth mentioning that, efficient adaptive beam tracking algorithms need to be employed for both BS and UE ends so that the two beams from BS and UE can be precisely aligned with acceptable latency.\\

 \item From the wireless hardware design point of view, the distributed phased arrays based architecture can help heat dissipation which is largely contributed by the PAs. In the state-of-the-art PA design for 5G phased arrays, the power added efficiency (PAE) is below 20\% \cite{Shakib:A 28}. Therefore the majority of the DC power will be converted into the thermal energy which increases the inner temperature of a mobile handset and potentially leads to a critical failure of the entire system. This issue is more pronounced when multiple mmWave PAs are integrated in the BF modules and the handset is operated at cell edge or with heavy traffic load. By arranging the mmWave BF modules in a distributed manner, it can largely mitigate this self-heating issue. Otherwise a cooling device is required \cite{Zihir:A 60}, however it is difficult to implement in a compact mobile handset.
\end{enumerate} 
 
\begin{figure*}
\centering
\includegraphics[scale = 1.03]{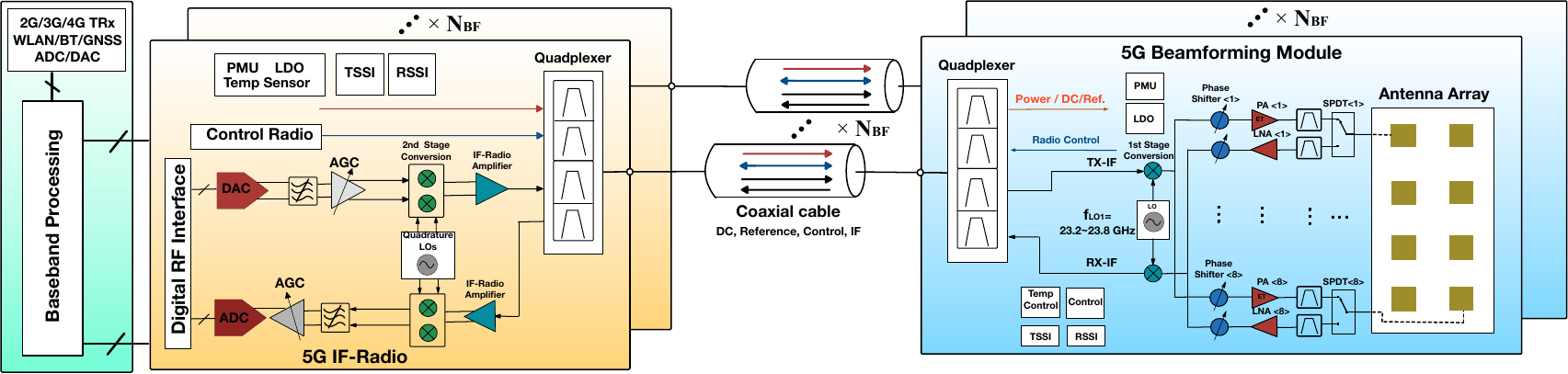}
\caption{Block diagram of 5G user equipment wireless system architecture.}\label{fig:5GBlockFig}
\end{figure*}
\subsection{Beamforming Module Hardware Design}
The details of a BF module design are given in Fig.~\ref{fig:BF_Fig}. First, the antenna array top-down view is shown in Fig.~\ref{fig:BF_Fig}(a). Second, the layer on which the chipsets are mounted is shown in Fig.~\ref{fig:BF_Fig}(b). A type of material with low dielectric constant and small loss tangent is desired. As a matter of fact, there are several suitable integration technology candidates such as low temperature co-fired ceramics (LTCC), hybrid LTCC \cite{Jin:60GHz}, multi-layer organics (MLO) \cite{Natarajan:A Fully}, liquid crystal polymer (LCP) \cite{Li:A Fully}, etc. Considering the cost, mass production and industrial maturity \cite{Dussopt:Silicon}, an MLO-like structure is adopted as it has shown profound value on commercial mass production in IEEE 802.11ad products \cite{Boers:60 GHz}.

In the cross section view of the BF module in Fig.~\ref{fig:BF_Fig}(c), Rogers RO4003C material is used for both antenna and PCB substrate because it has a low loss tangent and a suitable dielectric constant at the high GHz frequency. Accordingly, the effective wavelength ${\lambda }_{e}$ for the 28 GHz carrier can be calculated using (\ref{eq:WAVE})-(\ref{eq:DiEff2}). Moreover, the spacing among antenna elements is set to ${\lambda }_{0}$/2, thus $W$ and $L$ of the BF module are calculated as 25 and 18 mm, respectively, with some dimension margin. 

Each BF module is connected with the MLB by coaxial cables and coaxial connectors on PCBs, and BF modules are arranged in the back housing of the mobile devices. There are several critical factors in BF module arrangements. First, the spacing among BF modules, $d_\text{ISOL}$, should be kept sufficiently large ($\geq 1.5{\lambda }_{0}$, or 16 mm). Second, as long as a good spacing is guaranteed, more BF modules can be embedded on the back housing of the mobile device thus higher order MIMO can be obtained. However, this involves trade-offs between wireless performance and limited hardware area or resource on mobile devices. These constraints are much alleviated on tablet computers. The mmWave front-end, control and calibration circuits can be designed and fabricated using various conventional IC processes according to different design specifications and features. Connection between the PAs, LNAs and antenna elements is built by the stud bumps and the antenna feed lines routed inside the package. The through package vias (TPVs) route the signals between the dies and the PCB, and they also dissipate the heat which is mainly generated from the PA dies in the BF module. The thermal design and TPVs design are very critical because they are directly relevant to the cooling performance. A 5G BF module accommodates multiple PAs in a very small space, and a poor
thermal design may lead to its temperature rising to a threshold value and cause a total failure of the BF module. Moreover, an overheated BF module can result in a failure of the entire UE system, and in some extreme cases, it can even compromise the users' safety.
 
Normally, more than one IC process is used to implement the mmWave RF transceiver chipsets, and therefore, more than one die is used and shown in Fig.~\ref{fig:BF_Fig}(c). With the current mainstream IC processes, based on the emerging commercial products and the existing IC design experience, the total area of chipsets can be well managed below $10 \times 10$ mm according to mmWave IC designs in \cite{Boers:60 GHz}, \cite{Natarajan:A Fully} and \cite{Pang:60GHz}. Moreover, the thickness of the BF module can be briefly calculated using the equation as below
\begin{equation}\label{eq:thickness}
H_{\text{BF}}=H_{\text{ANT}}+H_{\text{bump}}+H_{\text{die}}+H_{\text{PCB}}+H_{\text{connector}}
\end{equation}
where $H_{\text{ANT}}$ is the thickness of the mmWave patch antenna with an estimated thickness of 0.4 mm, $H_{\text{bump}}$ stands for the stand bump thickness with a typical value of $\text{50 }\!\!\mu\!\!\text{ m}$, the die thickness $H_{\text{die}}$ is usually $\text{254 }\!\!\mu\!\!\text{ m}$, the thickness of PCB, $H_{\text{PCB}}$, is less than 0.4 mm, and the flat coaxial connector (widely used in commercial products) has a thickness of only 0.4 mm. Therefore, the total thickness including the surface mounted coaxial connector can be made below 1.5 mm, and consequently, a good form factor of 5G UE can be well maintained. 

Moreover, the patch antenna design can be flexible and tailored to specific requirement. For example, it can be a rectangular/circular microstrip patch, slot loop, Yagi-Uda, planar inverted-F, substrate integrated waveguide (SIW), etc. As long as they fit into the BF module to construct a phased array, they can be used in our DPA-MIMO architecture. The frequency band can be 28, 37, or 39 GHz for licensed 5G cellular networks.

\begin{figure*}
\centering
\includegraphics[scale = 0.8]{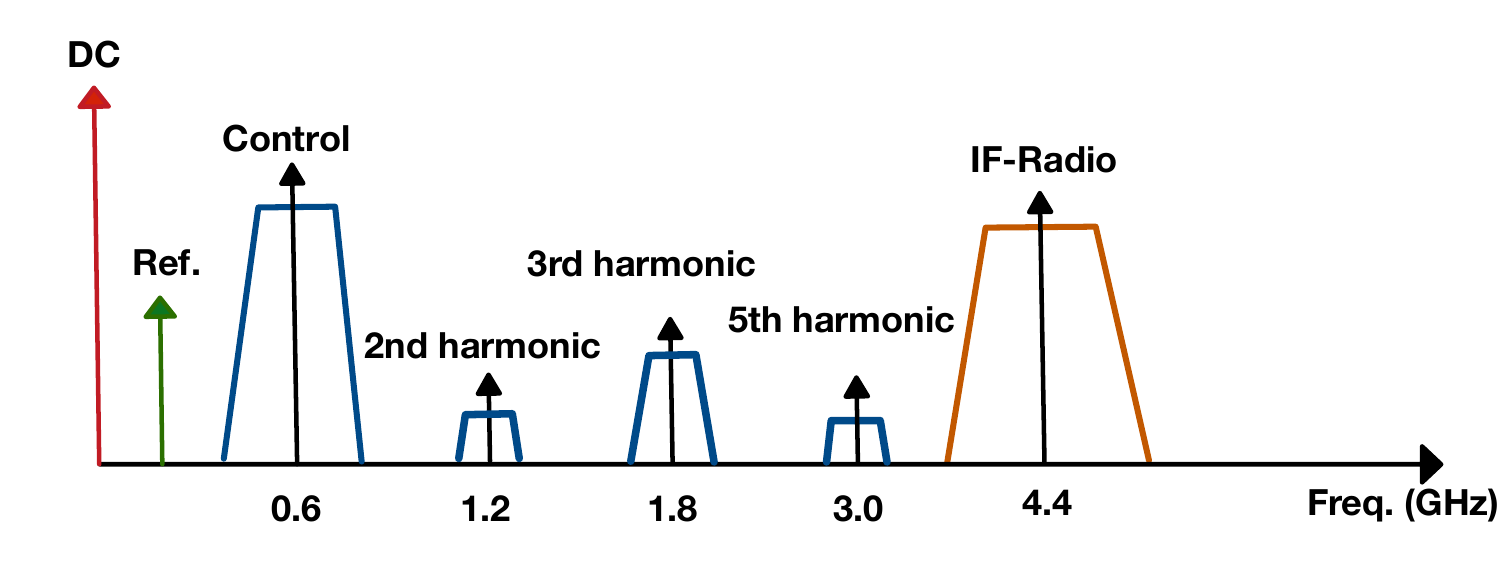}
\caption{5G user equipment wireless system frequency plan.}\label{fig:FrequencyPlanFig}
\end{figure*}

\subsection{RF Circuit Design of 5G Cellular UE}
Another important feature in the proposed RF circuit design is a split-IF architecture whose block diagram is depicted in Fig.~\ref{fig:5GBlockFig}. The BF module not only integrates the active antenna array to realize beamforming, but also enables frequency conversion for both uplink and downlink. As shown in Fig.~\ref{fig:5GBlockFig} the BF module down-converts a high GHz 5G band signal to a low GHz intermediate frequency (IF) signal in the downlink path, and up-converts an IF signal to a 5G band signal in the uplink path. The local oscillator (LO) signal $f_\text{LO1}$ can be tuned and realize the frequency conversion for a 28 GHz frequency band 5G TDD signal. Furthermore, this RF architecture can apply to 37 GHz, 39 GHz, and other 5G frequency bands after changing the LO frequency and corresponding hardware characteristics.     	

The motivation and benefits of utilizing this RF architecture lie in several aspects:
\begin{itemize}
 \item \emph{\textbf{First}}, the challenging requirement for the slim form factor of a contemporary handset design strictly limits the hardware dimension, and therefore it is not feasible to embed all BF modules on a larger MLB. \\ 
 \item \emph{\textbf{Second}}, the high-performance mmWave circuits design necessitates mmWave-enabled PCB such as Rogers RO4003C which is more costly but electrically less lossy than the FR-4 laminate widely used for the MLB design in contemporary smartphones. Therefore, separating BF modules and MLB design leads to cost effective manufacturing for mass production.\\ 
 \item \emph{\textbf{Third}}, converting the mmWave frequency to the IF frequency directly and immediately on a BF module minimizes the front-end insertion loss. Moreover, better signal integrity can be achieved since the connection is through coaxial cables instead of routing traces on the MLB. \\
 \item \emph{\textbf{Fourth}}, it offers the flexibility to handle various applications and scenarios without the need to reconfigure the entire wireless system design. For example, the placement and number of BF modules in the handset can be adjusted according to different system specifications and use cases, which makes it cost-effective.\\ 
 \item \emph{\textbf{Last}}, low GHz IF radio is less challenging to implement than its mmWave counterpart and therefore can be co-designed and manufactured in the same IC processes and SoCs for legacy cellular standards such as 3G and 4G. In addition, it can facilitate the test in mass production for both BF modules and IF radio plus baseband modules \cite{Boers:60 GHz}. 
\end{itemize}
    
Furthermore, a frequency plan is proposed as follows. Take the 28 GHz frequency band as an example. As shown in Fig.~\ref{fig:FrequencyPlanFig}, the IF frequency is set to 4.4 GHz, the radio control signal is at 600 MHz, the reference clock signal is set to below 100 MHz, and power supply is a DC signal. These signals are all supplied over the coaxial cable, and separated or combined using quadplexers on both the BF modules and IF radio ends. The IF frequency is chosen at 4.4 GHz due to several reasons. First, it does not fall in any LTE band of 3GPP Rel. 14, neither any WiFi/GNSS frequency so that the desense issue of LTE/WiFi can be mitigated; second, when conducting frequency up-conversion, its image frequency can be easily filtered out since the image frequency is separated from the desired frequency by 8.8 GHz; third, the IF frequency is in the low GHz range and therefore using cost-effective coaxial cables can satisfy the performance requirement. 

In addition, the control signal is chosen to be operated at 600 MHz. As shown in Fig.~\ref{fig:FrequencyPlanFig}, the harmonics of the control signal do not interfere with the IF radio signal, and hence the desense issue can be mitigated. A control interface provides supervision and operation of BF modules through read-write to the registers using radio control signals such as RF front-end (RFFE) control interface signals. The RFFE signals carry the information of transmitter signal strength indicator (TSSI), receiver signal strength indicator (RSSI), and it executes the calibration and temperature control of a BF module. 
\begin{figure}
\centering
\includegraphics[scale = 0.55]{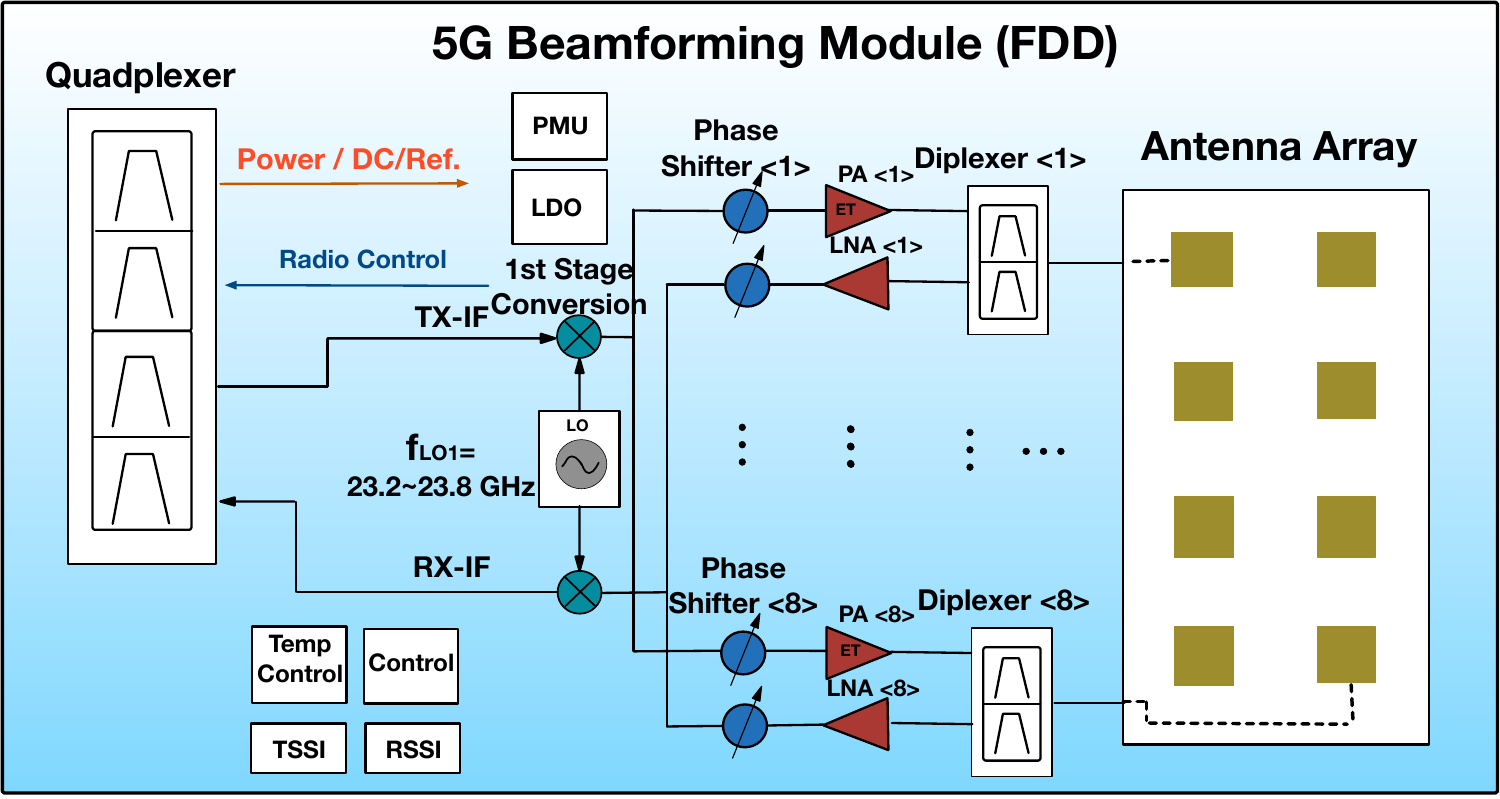}
\caption{Block diagram of 5G beamforming module supporting FDD duplex scheme.}\label{fig:FDDFig}
\end{figure}  

Assume the 28 GHz band contains four sub-bands with each one occupying a bandwidth of 200 MHz and some guard bands. The BF module can support a wide bandwidth up to 800 MHz which is equivalent to four intra sub-bands aggregated. The local oscillator in the BF module should cover a frequency range at least from 23.2 to 23.9 GHz. Therefore, a 4.4 GHz IF radio signal can be obtained after doing frequency down-conversion, and a 28 GHz 5G mmWave signal can be generated after conducting frequency up-conversion. Totally, $N_\text{BF}$ BF modules and $N_\text{BF}$ IF radios are integrated to support a maximum of $N_\text{BF}$ streams communicating simultaneously. 

As to the detailed BF module implementations, PMUs and low drop-out regulators (LDOs) in each 5G BF module transform the DC voltage of the coax cable to different power supplies for different dies. At the RF front-end, each antenna element is connected to one single port double throw (SPDT) which is controlled by a radio control signal to enable time domain duplexing. PAs and low noise amplifiers (LNAs) are respectively placed in the uplink and downlink paths followed by digitally controlled phase shifters (DCPS) which determine the step resolution of beam steering. For the IF radio module design, the direct conversion RF architecture is employed. Moreover, the differential to single-ended (D-to-S) amplifier and the single-ended to differential (S-to-D) converter are situated in the uplink and downlink paths respectively. The variable gain amplifier (VGA) realizes the function of automatic gain control (AGC) so that the dynamic range (DR) requirement of the analog-to-digital converter (ADC) can be mitigated. On the other hand, for a FDD based 5G beamforming module, the SPDT and filters in the TDD mode are replaced with the diplexers as depicted in Fig.~\ref{fig:FDDFig}.

\subsection{Advancement of Data Converter Techniques}
With respect to the ADCs and digital-to-analog converters (DACs), they should support a wide RF bandwidth with high resolution which depends on the actual application, for example the order of digital modulation and the performance of VGA. In this proposed 5G cellular UE, 256-QAM is supported, which needs a resolution of 12 bits or above. Furthermore, a high spur free dynamic range of ADC needs to be maintained considering that the input signal at the receiver end can range from around -25 dBm to -110 dBm \cite{Anzaldo:Navigate}. The high dynamic range requirement of the wideband VGA can be alleviated by using high-performance data converters. 

There has been a concern of the high power consumption and poor cost effectiveness of ADC for 5G applications \cite{Liu:The}. In fact, several state-of-the-art designs have recently demonstrated satisfying performance such as low power consumption and small chip area. In \cite{Nam:A 12}, a 12-bit, 1.6 GS/s time interleaved ADC only consumes 37.7 mW with 0.9 ${\text{mm}^2}$ chip area, and it achieves 17.8 fJ/conversion. In other words, such ADC can enable a theoretical absolute physical data throughput of 10.8 Gb/s for an ideal 256-QAM demodulation. Moreover, \cite{Lin:A 10}-\cite{Su:A 12} have presented high-performance, energy and area efficient ADC and DAC which can be considered as good prototype candidates for future 5G UE data converters. Schreier figure of merit ($\text{FOM}_{\text{S}}$) and Walden figure of merit ($\text{FOM}_{\text{W}}$) \cite{Walden:ADC} are commonly used to evaluate the data converters performance, as expressed below
\begin{equation}\label{eq:FOMADC1}
\begin{aligned}
\ \text{FOM}_{\text{S}}=\text{SNDR}+10\text{log}_\text{10}(B/P)
\end{aligned}
\end{equation}
\begin{equation}\label{eq:FOMADC2}
\begin{aligned}
\ \text{FOM}_{\text{W}}=P/(2^{\text{ENOB} }\times{\text{min}(2B,f_{s})}), \
\end{aligned}
\end{equation}
where $\text{SNDR}$ is the signal-to-noise-and-distortion ratio,  $\text{ENOB}$ stands for the effective number of bits, $B$ is the analog bandwidth, $P$ is the power consumption, and ${f}_\text{S}$ is the sampling rate. According to the data collected \cite{Murmann:ADC} which has summarized the performance of state-of-the-art ADCs published in the International Solid-State Circuits Conference (ISSCC) and the Symposia on VLSI Technology and Circuits (VLSI Symposia) in recent 20 years, there are a couple of designs suitable for 200 MHz wide or even 800 MHz wide analog frequency, with a $\text{FOM}_{\text{W}}$ smaller than 50 fJ/conversion. Moreover, the FOM of data converters keeps improving at a steady pace which will further facilitate the 5G UE hardware design.
\begin{figure}[!h]
\centering
\includegraphics[scale = 0.45]{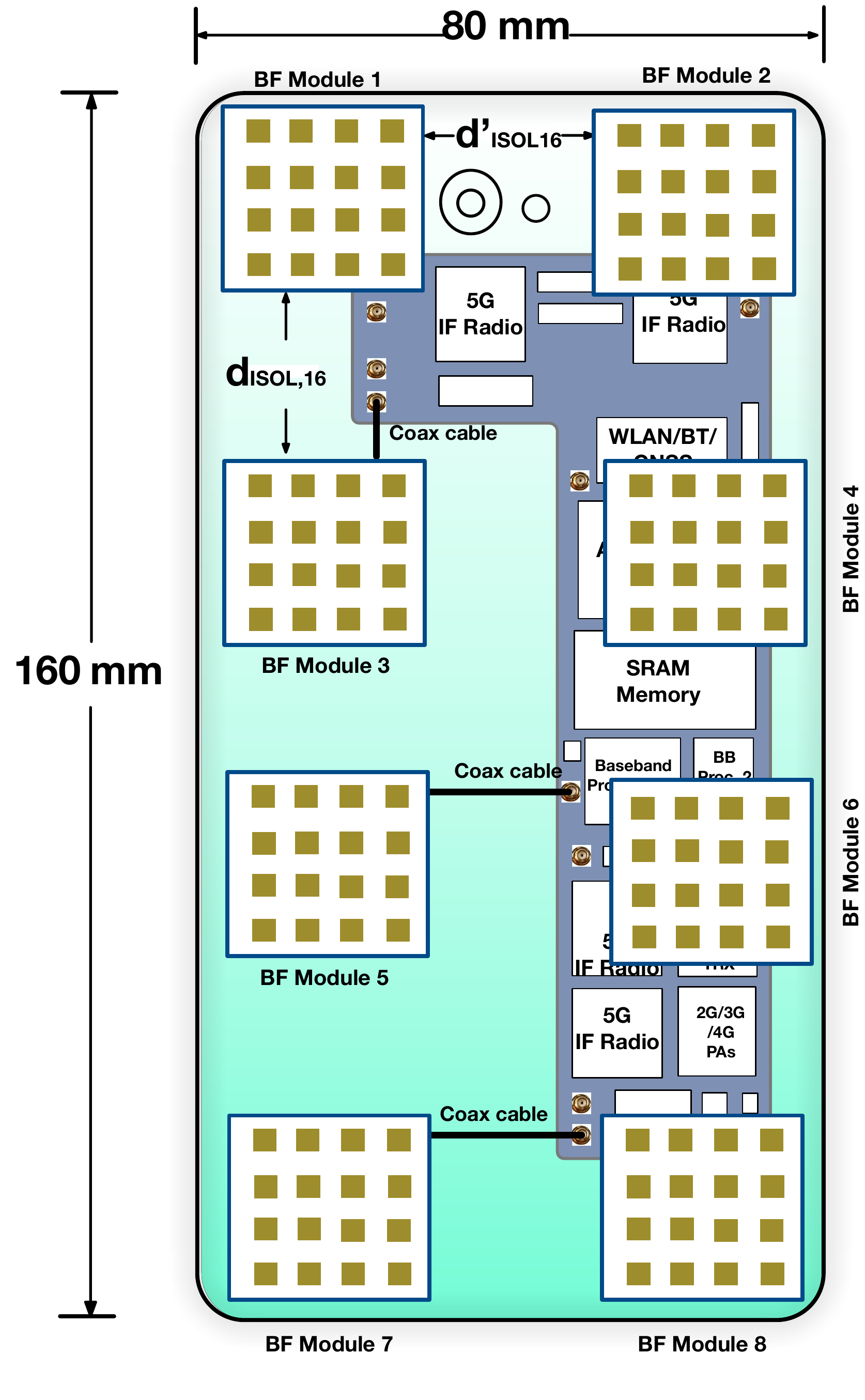}
\caption{DPA-MIMO system in a mobile phone handset when $N_\text{ANT}$=16. }\label{fig:5GUE16Fig}
\end{figure}

\begin{table*}[!h]  % DOWNLINK
\scriptsize
\caption{Calculation and comparison of downlink budget at 28 GHz.} \label{tab:PDLT}
\newcommand{\tabincell}[2]{\begin{tabular}{@{}#1@{}}#2\end{tabular}}
\centering
 \begin{threeparttable}
 
\newcolumntype{C}{>{\centering\arraybackslash}p{2em}} 
  \begin{tabular}{|c|c|c|c|c|c|c|c|c|c|c|c|c|c|c|c|c|}\hline
 
        \tabincell{c}{} & \multicolumn{16}{c|}{\textbf{Popular deployment scenarios}}      \\  \hline       
        \tabincell{c}{\textbf{5G cellular}\\\textbf{service link}\\\textbf{budget}} & \multicolumn{2}{c|}{\tabincell{c}{UMi-street\\open-NLOS\\d=100 m}} & \multicolumn{2}{c|}{\tabincell{c}{UMi-street\\open-NLOS\\d=200 m}} & \multicolumn{2}{c|}{\tabincell{c}{UMi-street\\canyon-NLOS\\d=100 m}} & \multicolumn{2}{c|}{\tabincell{c}{ UMi-street\\canyon-NLOS\\d=200 m}} & \multicolumn{2}{c|}{\tabincell{c}{ UMa-NLOS\\d=200 m}} & \multicolumn{2}{c|}{\tabincell{c}{UMa-NLOS\\d=500 m}} & \multicolumn{2}{c|}{\tabincell{c}{UMa-NLOS\\d=1000 m}} & \multicolumn{2}{c|}{\tabincell{c}{UMa-NLOS\\d=2000 m}}      \\  \hline
        
        \tabincell{c}{Bandwidth \\(MHz)} & \multicolumn{2}{c|}{\tabincell{c}{200}} & \multicolumn{2}{c|}{\tabincell{c}{200}} & \multicolumn{2}{c|}{\tabincell{c}{200}} & \multicolumn{2}{c|}{\tabincell{c}{200}} & \multicolumn{2}{c|}{\tabincell{c}{200}} & \multicolumn{2}{c|}{\tabincell{c}{200}} & \multicolumn{2}{c|}{\tabincell{c}{200}} & \multicolumn{2}{c|}{\tabincell{c}{200}}      \\  \hline
        
        \tabincell{c}{Max EIRP \\(dBm)} & \multicolumn{2}{c|}{\tabincell{c}{78}} & \multicolumn{2}{c|}{\tabincell{c}{78}} & \multicolumn{2}{c|}{\tabincell{c}{78}} & \multicolumn{2}{c|}{\tabincell{c}{78}} & \multicolumn{2}{c|}{\tabincell{c}{78}} & \multicolumn{2}{c|}{\tabincell{c}{78}} & \multicolumn{2}{c|}{\tabincell{c}{78}} & \multicolumn{2}{c|}{\tabincell{c}{78}}      \\  \hline
        
        \tabincell{c}{Path loss (dB)} & \multicolumn{2}{c|}{\tabincell{c}{126.3}} & \multicolumn{2}{c|}{\tabincell{c}{135.0}} & \multicolumn{2}{c|}{\tabincell{c}{133.4}} & \multicolumn{2}{c|}{\tabincell{c}{143.0}} & \multicolumn{2}{c|}{\tabincell{c}{137.2}} & \multicolumn{2}{c|}{\tabincell{c}{149.2}} & \multicolumn{2}{c|}{\tabincell{c}{158.2}} & \multicolumn{2}{c|}{\tabincell{c}{167.2}}      \\  \hline

        \tabincell{c}{Received \\power (dBm)} & \multicolumn{2}{c|}{\tabincell{c}{-48.3}} & \multicolumn{2}{c|}{\tabincell{c}{-57.0}} & \multicolumn{2}{c|}{\tabincell{c}{-55.4}} & \multicolumn{2}{c|}{\tabincell{c}{-65.0}} & \multicolumn{2}{c|}{\tabincell{c}{-59.2}} & \multicolumn{2}{c|}{\tabincell{c}{-71.2}} & \multicolumn{2}{c|}{\tabincell{c}{-80.2}} & \multicolumn{2}{c|}{\tabincell{c}{-89.2}}      \\  \hline        

         \tabincell{c}{Thermal \\noise (dBm)} & \multicolumn{2}{c|}{\tabincell{c}{-91.0}} & \multicolumn{2}{c|}{\tabincell{c}{-91.0}} & \multicolumn{2}{c|}{\tabincell{c}{-91.0}} & \multicolumn{2}{c|}{\tabincell{c}{-91.0}} & \multicolumn{2}{c|}{\tabincell{c}{-91.0}} & \multicolumn{2}{c|}{\tabincell{c}{-91.0}} & \multicolumn{2}{c|}{\tabincell{c}{-91.0}} & \multicolumn{2}{c|}{\tabincell{c}{-91.0}}      \\  \hline           

         \tabincell{c}{SNR before\\BF (dB)} & \multicolumn{2}{c|}{\tabincell{c}{42.7}} & \multicolumn{2}{c|}{\tabincell{c}{34.0}} & \multicolumn{2}{c|}{\tabincell{c}{35.6}} & \multicolumn{2}{c|}{\tabincell{c}{20.0}} & \multicolumn{2}{c|}{\tabincell{c}{31.8}} & \multicolumn{2}{c|}{\tabincell{c}{19.8}} & \multicolumn{2}{c|}{\tabincell{c}{10.8}} & \multicolumn{2}{c|}{\tabincell{c}{1.8}}      \\  \hline 
         
\tabincell{c}{Rx front\\end loss\tnote{1} (dB)} & \multicolumn{2}{c|}{\tabincell{c}{4.0}} & \multicolumn{2}{c|}{\tabincell{c}{4.0}} & \multicolumn{2}{c|}{\tabincell{c}{4.0}} & \multicolumn{2}{c|}{\tabincell{c}{4.0}} & \multicolumn{2}{c|}{\tabincell{c}{4.0}} & \multicolumn{2}{c|}{\tabincell{c}{4.0}} & \multicolumn{2}{c|}{\tabincell{c}{4.0}} & \multicolumn{2}{c|}{\tabincell{c}{4.0}}      \\  \hline                     
         
         \tabincell{c}{Single antenna\\element gain (dB)} & \multicolumn{2}{c|}{\tabincell{c}{5.0}} & \multicolumn{2}{c|}{\tabincell{c}{5.0}} & \multicolumn{2}{c|}{\tabincell{c}{5.0}} & \multicolumn{2}{c|}{\tabincell{c}{5.0}} & \multicolumn{2}{c|}{\tabincell{c}{5.0}} & \multicolumn{2}{c|}{\tabincell{c}{5.0}} & \multicolumn{2}{c|}{\tabincell{c}{5.0}} & \multicolumn{2}{c|}{\tabincell{c}{5.0}}      \\  \hline 
          \tabincell{c}{\textbf{$N_\text{ANT}$, number of antenna}\\ \textbf{elements in each BF}} & \textbf{8} & \textbf{16} & \textbf{8} & \textbf{16} & \textbf{8} & \textbf{16} & \textbf{8} & \textbf{16} & \textbf{8} & \textbf{16} & \textbf{8} & \textbf{16} & \textbf{8} & \textbf{16} & \textbf{8} & \textbf{16}   \\ \hline 
                 
          \tabincell{c}{\text{Total antenna}\\ \text{array gain (dB)}} & 14 & 17 & 14 & 17 & 14 & 17 & 14 & 17 & 14 & 17 & 14 & 17 & 14 & 17 & 14 & 17   \\ \hline   

          \tabincell{c}{\text{Noise figure}\\ \text{(dB)}} & 7.0 & 7.0 & 7.0 & 7.0 & 7.0 & 7.0 & 7.0 & 7.0 & 7.0 & 7.0 & 7.0 & 7.0 & 7.0 & 7.0 & 7.0 & 7.0   \\ \hline 
         
          \tabincell{c}{SNR after\\BF (dB)} & 45.7 & 48.7 & 37.0 & 40.0 & 38.6 & 41.6 & 29.0 & 32.0 & 34.8 & 37.8 & 22.8 & 25.8 & 13.8 & 16.8 & 4.8 & 7.8   \\ \hline 
          
          \tabincell{c}{\text{Spectral efficiency}\\ \text{SISO\tnote{2} (bits/s/Hz)}} & \multicolumn{2}{c|}{\tabincell{c}{8}} & \multicolumn{2}{c|}{\tabincell{c}{8}} & \multicolumn{2}{c|}{\tabincell{c}{8}} & \tabincell{c}{7.98} & \tabincell{c}{8} & \multicolumn{2}{c|}{\tabincell{c}{8}} & \tabincell{c}{7.11} & \tabincell{c}{7.76} & \tabincell{c}{4.35} & \tabincell{c}{5.18} & \tabincell{c}{1.69} & \tabincell{c}{1.94}      \\  \hline

          \tabincell{c}{\text{BW=200 MHz} \\ \text{SISO\tnote{3}} \\ \text{throughput (Mbps)}} & \multicolumn{2}{c|}{\tabincell{c}{1280}} & \multicolumn{2}{c|}{\tabincell{c}{1280}} & \multicolumn{2}{c|}{\tabincell{c}{1280}} & \tabincell{c}{1280} & \tabincell{c}{1280} & \multicolumn{2}{c|}{\tabincell{c}{1280}} & \tabincell{c}{1138} & \tabincell{c}{1242} & \tabincell{c}{696} & \tabincell{c}{828} & \tabincell{c}{270} & \tabincell{c}{310}      \\  \hline  
         
          \tabincell{c}{\text{BW=200 MHz}\\ \text{$8\times8$ MIMO} \\ \text{throughput (Mbps)}} & \multicolumn{2}{c|}{\tabincell{c}{10240}} & \multicolumn{2}{c|}{\tabincell{c}{10240}} & \multicolumn{2}{c|}{\tabincell{c}{10240}} & \tabincell{c}{10240} & \tabincell{c}{10240} & \multicolumn{2}{c|}{\tabincell{c}{10240}} & \tabincell{c}{9104} & \tabincell{c}{9936} & \tabincell{c}{5568} & \tabincell{c}{6624} & \tabincell{c}{2160} & \tabincell{c}{2480}      \\  \hline 
          
          \tabincell{c}{\text{BW=800 MHz}\\ \text{SISO} \\ \text{throughput (Mbps)}} & \multicolumn{2}{c|}{\tabincell{c}{5120}} & \multicolumn{2}{c|}{\tabincell{c}{5120}} & \multicolumn{2}{c|}{\tabincell{c}{5120}} & \tabincell{c}{5120} & \tabincell{c}{5120} & \multicolumn{2}{c|}{\tabincell{c}{5120}} & \tabincell{c}{4552} & \tabincell{c}{4968} & \tabincell{c}{2784} & \tabincell{c}{3312} & \tabincell{c}{1080} & \tabincell{c}{1240}      \\  \hline 
          
          \tabincell{c}{\text{BW=800 MHz}\\ \text{$8\times8$ MIMO}\\ \text{throughput (Mbps)}} & \multicolumn{2}{c|}{\tabincell{c}{40960}} & \multicolumn{2}{c|}{\tabincell{c}{40960}} & \multicolumn{2}{c|}{\tabincell{c}{40960}} & \tabincell{c}{40960} & \tabincell{c}{40960} & \multicolumn{2}{c|}{\tabincell{c}{40960}} & \tabincell{c}{36416} & \tabincell{c}{39744} & \tabincell{c}{22272} & \tabincell{c}{26496} & \tabincell{c}{8640} & \tabincell{c}{9920}      \\  \hline      
   
    \end{tabular}
    \begin{tablenotes}
        \footnotesize
        \item[1] Before the LNA stage of the BF, the front-end loss includes insertion loss of signal traces, switches and filters
        \item[2] Based on 256-QAM modulation
        \item[3] Based on 20\% system overhead which is a typical case in LTE \cite{Agilent:4G}, \cite{Chris:LTE}
      \end{tablenotes}
    \end{threeparttable}
\end{table*}

\section{LINK BUDGET CALCULATION AND WIRELESS PERFORMANCE EVALUATION}
In this section, link budget analysis and throughput estimation are conducted for the downlink and uplink of the proposed DPA-MIMO UE design. The 5G channel model in \cite{NYU:5G} is used, and the numbers used for insertion loss, noise figure (NF) and antenna gain are explained first. The performance of a mmWave antenna switch in \cite{Byeon:Design} shows that IL can be well managed below 1.9 dB with a TX-RX isolation better than 38 dB. On the other hand, according to the state-of-the-art band pass filter (BPF) design \cite{Yang:A 20}, the IL is below 1.5 dB. Therefore, in Table~\ref{tab:PDLT} and Table~\ref{tab:PULT}, the RX front-end loss before the LNA is set to 4.0 dB including extra loss due to the interface between the LNA and the antenna elements. The noise figure of the mmWave receiver varies with different IC processes, and for the state-of-the-art CMOS designs in \cite{Boers:60 GHz} and \cite{Saito:A Fully}, NF is 7.1 dB and 8 dB, respectively. For the design using more advanced IC process such as SiGe, BiCMOS, NF of a receiver can achieve 6.8 dB \cite{Natarajan:A Fully} and 5.5 dB \cite{Tomkins:A 60}. Considering the superior cost-effectiveness of the CMOS process and its widespread use, it is reasonable to assume that 5G receiver NF is around 7 dB. Furthermore, the calculated antenna gain of one single patch antenna element can achieve 5 to 7 dBi in state-of-the-art designs \cite{Zihir:A 60}, \cite{Jin:60GHz}. Therefore in the following tables, single antenna element gain is set to 5 dBi.

\begin{figure*}[!h]
\centering
\includegraphics[scale = 0.7]{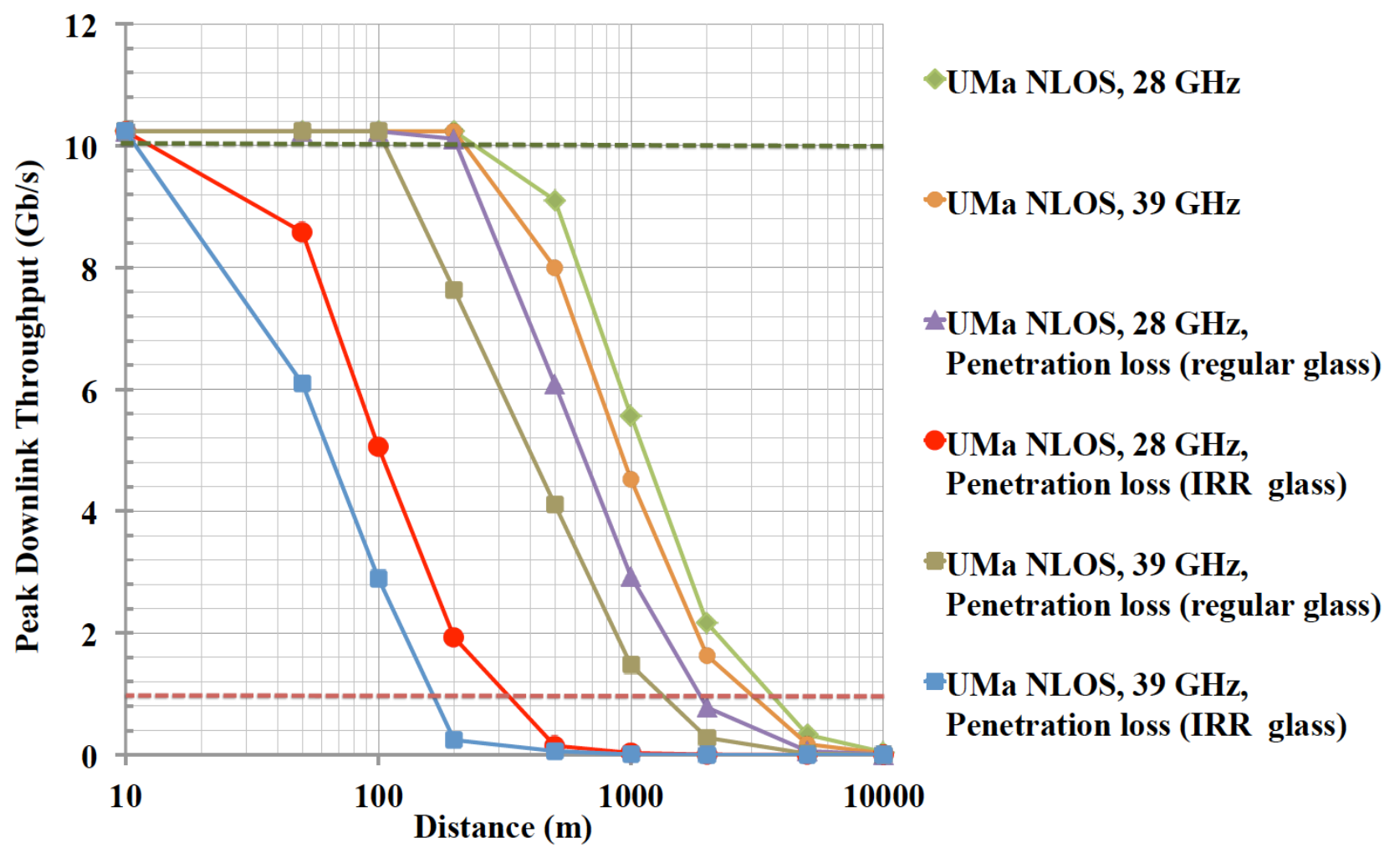}
\caption{$8\times8$ MIMO, BW=200 MHz, $N_\text{ue}$=$8\times8$, peak downlink throughput versus various deployment scenarios.}\label{fig:PDLTFig}
\end{figure*}

\begin{figure}[!h]
\centering
\includegraphics[scale = 0.5]{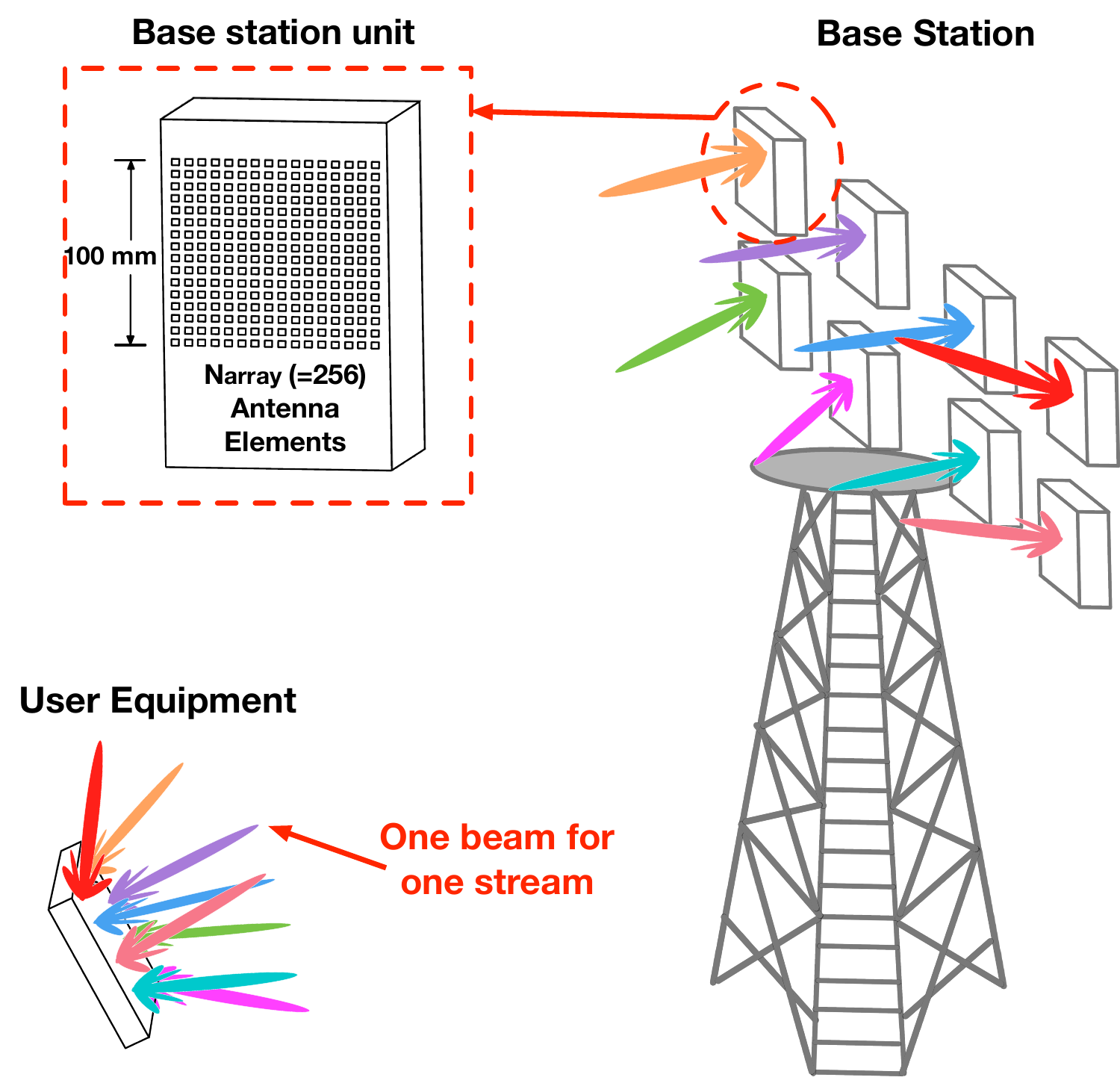}
\caption{Communication between user equipment and base station in $8\times8$ MIMO mode.}\label{fig:BSFig}
\end{figure}

\begin{table*}[!h]   % UPLINK
\scriptsize
\caption{Calculation and comparison of uplink budget at 28 GHz.} \label{tab:PULT}
\newcommand{\tabincell}[2]{\begin{tabular}{@{}#1@{}}#2\end{tabular}}
\centering
% \begin{threeparttable}
 
\newcolumntype{C}{>{\centering\arraybackslash}p{2em}} 

  \begin{tabular}{|c|c|c|c|c|c|c|c|c|c|c|c|c|c|c|c|c|}\hline
 
        \tabincell{c}{} & \multicolumn{16}{c|}{\textbf{Popular deployment scenarios}}      \\  \hline     
          
        \tabincell{c}{\textbf{5G cellular}\\\textbf{service link}\\\textbf{budget}} & \multicolumn{2}{c|}{\tabincell{c}{UMi-street\\open-NLOS\\d=100 m}} & \multicolumn{2}{c|}{\tabincell{c}{UMi-street\\open-NLOS\\d=200 m}} & \multicolumn{2}{c|}{\tabincell{c}{UMi-street\\canyon-NLOS\\d=100 m}} & \multicolumn{2}{c|}{\tabincell{c}{ UMi-street\\canyon-NLOS\\d=200 m}} & \multicolumn{2}{c|}{\tabincell{c}{ UMa-NLOS\\d=200 m}} & \multicolumn{2}{c|}{\tabincell{c}{UMa-NLOS\\d=500 m}} & \multicolumn{2}{c|}{\tabincell{c}{UMa-NLOS\\d=1000 m}} & \multicolumn{2}{c|}{\tabincell{c}{UMa-NLOS\\d=2000 m}}      \\  \hline
        
        \tabincell{c}{Bandwidth \\(MHz)} & \multicolumn{2}{c|}{\tabincell{c}{200}} & \multicolumn{2}{c|}{\tabincell{c}{200}} & \multicolumn{2}{c|}{\tabincell{c}{200}} & \multicolumn{2}{c|}{\tabincell{c}{200}} & \multicolumn{2}{c|}{\tabincell{c}{200}} & \multicolumn{2}{c|}{\tabincell{c}{200}} & \multicolumn{2}{c|}{\tabincell{c}{200}} & \multicolumn{2}{c|}{\tabincell{c}{200}}      \\  \hline
        
        \tabincell{c}{Max EIRP \\(dBm)} & \multicolumn{2}{c|}{\tabincell{c}{43}} & \multicolumn{2}{c|}{\tabincell{c}{43}} & \multicolumn{2}{c|}{\tabincell{c}{43}} & \multicolumn{2}{c|}{\tabincell{c}{43}} & \multicolumn{2}{c|}{\tabincell{c}{43}} & \multicolumn{2}{c|}{\tabincell{c}{43}} & \multicolumn{2}{c|}{\tabincell{c}{43}} & \multicolumn{2}{c|}{\tabincell{c}{43}}      \\  \hline
        
        \tabincell{c}{Path loss (dB)} & \multicolumn{2}{c|}{\tabincell{c}{126.3}} & \multicolumn{2}{c|}{\tabincell{c}{135.0}} & \multicolumn{2}{c|}{\tabincell{c}{133.4}} & \multicolumn{2}{c|}{\tabincell{c}{143.0}} & \multicolumn{2}{c|}{\tabincell{c}{137.2}} & \multicolumn{2}{c|}{\tabincell{c}{149.2}} & \multicolumn{2}{c|}{\tabincell{c}{158.2}} & \multicolumn{2}{c|}{\tabincell{c}{167.2}}      \\  \hline

        \tabincell{c}{Received \\power (dBm)} & \multicolumn{2}{c|}{\tabincell{c}{-83.3}} & \multicolumn{2}{c|}{\tabincell{c}{-92.0}} & \multicolumn{2}{c|}{\tabincell{c}{-90.4}} & \multicolumn{2}{c|}{\tabincell{c}{-100.0}} & \multicolumn{2}{c|}{\tabincell{c}{-94.2}} & \multicolumn{2}{c|}{\tabincell{c}{-106.2}} & \multicolumn{2}{c|}{\tabincell{c}{-115.2}} & \multicolumn{2}{c|}{\tabincell{c}{-124.2}}      \\  \hline        

         \tabincell{c}{Thermal \\noise (dBm)} & \multicolumn{2}{c|}{\tabincell{c}{-91.0}} & \multicolumn{2}{c|}{\tabincell{c}{-91.0}} & \multicolumn{2}{c|}{\tabincell{c}{-91.0}} & \multicolumn{2}{c|}{\tabincell{c}{-91.0}} & \multicolumn{2}{c|}{\tabincell{c}{-91.0}} & \multicolumn{2}{c|}{\tabincell{c}{-91.0}} & \multicolumn{2}{c|}{\tabincell{c}{-91.0}} & \multicolumn{2}{c|}{\tabincell{c}{-91.0}}      \\  \hline           

         \tabincell{c}{SNR before\\BF (dB)} & \multicolumn{2}{c|}{\tabincell{c}{7.7}} & \multicolumn{2}{c|}{\tabincell{c}{-1.0}} & \multicolumn{2}{c|}{\tabincell{c}{0.6}} & \multicolumn{2}{c|}{\tabincell{c}{-9.0}} & \multicolumn{2}{c|}{\tabincell{c}{-3.2}} & \multicolumn{2}{c|}{\tabincell{c}{-15.2}} & \multicolumn{2}{c|}{\tabincell{c}{-24.2}} & \multicolumn{2}{c|}{\tabincell{c}{-33.2}}      \\  \hline 
         
\tabincell{c}{Rx front\\end loss (dB)} & \multicolumn{2}{c|}{\tabincell{c}{4.0}} & \multicolumn{2}{c|}{\tabincell{c}{4.0}} & \multicolumn{2}{c|}{\tabincell{c}{4.0}} & \multicolumn{2}{c|}{\tabincell{c}{4.0}} & \multicolumn{2}{c|}{\tabincell{c}{4.0}} & \multicolumn{2}{c|}{\tabincell{c}{4.0}} & \multicolumn{2}{c|}{\tabincell{c}{4.0}} & \multicolumn{2}{c|}{\tabincell{c}{4.0}}      \\  \hline                     
         
         \tabincell{c}{Single antenna\\element gain (dB)} & \multicolumn{2}{c|}{\tabincell{c}{5.0}} & \multicolumn{2}{c|}{\tabincell{c}{5.0}} & \multicolumn{2}{c|}{\tabincell{c}{5.0}} & \multicolumn{2}{c|}{\tabincell{c}{5.0}} & \multicolumn{2}{c|}{\tabincell{c}{5.0}} & \multicolumn{2}{c|}{\tabincell{c}{5.0}} & \multicolumn{2}{c|}{\tabincell{c}{5.0}} & \multicolumn{2}{c|}{\tabincell{c}{5.0}}      \\  \hline 
          \textbf{$N_\text{array}$} & \textbf{64} & \textbf{256} & \textbf{64} & \textbf{256} & \textbf{64} & \textbf{256} & \textbf{64} & \textbf{256} & \textbf{64} & \textbf{256} & \textbf{64} & \textbf{256} & \textbf{64} & \textbf{256} & \textbf{64} & \textbf{256}   \\ \hline 
                 
          \tabincell{c}{\text{Total antenna}\\ \text{array gain (dB)}} & 23 & 29 & 23 & 29 & 23 & 29 & 23 & 29 & 23 & 29 & 23 & 29 & 23 & 29 & 23 & 29   \\ \hline   

          \tabincell{c}{\text{Noise figure}\\ \text{(dB)}} & 7.0 & 7.0 & 7.0 & 7.0 & 7.0 & 7.0 & 7.0 & 7.0 & 7.0 & 7.0 & 7.0 & 7.0 & 7.0 & 7.0 & 7.0 & 7.0   \\ \hline 
         
          \tabincell{c}{SNR after\\BF (dB)} & 19.8 & 25.8 & 11.1 & 17.1 & 12.6 & 18.6 & 3.0 & 9.0 & 8.8 & 14.8 & -3.2 & 2.8 & -12.2 & -6.2 & -21.2 & -15.2   \\ \hline 
          
          \tabincell{c}{Spectral efficiency \\ \text{SISO (bits/s/Hz)}} & 6.19 & 7.75 &3.56 &5.25 & 4.01 &5.6 & 1.55 &2.95 &2.90 &4.64 &0.57 &1.52 &0.08 &0.31 & 0.01 & 0.04      \\  \hline       
           
          \tabincell{c}{BW=200 MHz\\ SISO\\ throughput (Mbps)} & 989 & 1240 & 570 & 839 & 642 & 896 & 248 & 472 & 464 & 742 & 92 & 244 & 13.7 & 50 & 1.7 & 7      \\  \hline  
         
          \tabincell{c}{Total antenna \\ elements of BS} & 512 & 2048 & 512 & 2048 & 512 & 2048 & 512 & 2048 & 512 & 2048 & 512 & 2048 & 512 & 2048 & 512 & 2048      \\  \hline           
         
          \tabincell{c}{BW=200 MHz \\ \text{$8\times8$ MIMO}\\ throughput (Mbps)} & 7912 & 9920 & 4560 & 6952 & 5136 & 7168 & 1984 & 3776 & 3712 & 5936 & 736 & 1952 & 109.6 & 400 & 13.6 & 56      \\  \hline 
          
          \tabincell{c}{BW=800 MHz \\ SISO\\ throughput (Mbps)} & 3956 & 4960 & 2280 & 3476 & 2568 & 3584 & 992 & 1888 & 1856 & 2968 & 368 & 976 & 54.86 & 200 & 6.8 & 28     \\  \hline 
          
          \tabincell{c}{\text{BW=800 MHz}\\ \text{$8\times8$ MIMO}\\ throughput (Mbps)} & 31648 & 39680 & 18240 & 27808 & 20544 & 28672 & 7936 & 15104 & 14848 & 23744 & 2944 & 7808 & 438.4 & 1600 & 54.4 & 224      \\  \hline      
   
    \end{tabular}
\end{table*}

\subsection{Downlink Budget and Data Throughput Analysis}
The downlink budget calculation under several popular deployment scenarios is given in Table~\ref{tab:PDLT}, also with the results from numerical analysis for the data throughput. The UMi Street Canyon NLOS and UMa NLOS scenarios in the new 5G channel model \cite{NYU:5G} have been chosen for their larger path loss and shadowing coefficients as the worst-case calculation. Typically, the maximum radius of a microcell is 200 meters, and a macrocell BS can cover up to more than 1 km. Therefore, the UMa NLOS model is used to represent communication distance more than 200 meters. 
    
As can be analyzed from the results in Table~\ref{tab:PDLT}, in some cases, SE becomes smaller as the SNR decreases because when the SNR goes down below some threshold value, a lower digital modulation order is enabled. Furthermore, two sets of data are given in Table~\ref{tab:PDLT} based on 8 and 16 antenna elements per BF module respectively. The 16 antenna elements based UE architecture is drawn in Fig.~\ref{fig:5GUE16Fig} which shows that a sufficiently large physical separation ($>2\lambda_\text{0}$ in Fig.~\ref{fig:5GUE16Fig}) is well maintained. By using more antenna elements, it can increase the receiver gain, boost the SNR and EIRP so that the mobile handset can operate in a more challenging environment and handle larger path loss and penetration loss of buildings, particularly for mmWave frequency bands. 

In Fig.~\ref{fig:PDLTFig}, the PDLT under several deployment scenarios is given for 28 GHz and 39 GHz respectively. The number of antenna elements in one UE, denoted as $N_\text{ue}$ (equals $N_\text{ANT} \times N_\text{BF}$), is set to $8\times8$ (=64). The bandwidth is set to 200 MHz, and the UE is configured in MIMO with a maximum of $N_\text{BF}$ (equals 8 in Fig.~\ref{fig:PDLTFig}) layers. The penetration loss models of regular glass and infrared reflective (IRR) glass presented in \cite{Haneda:5G} are also added into the propagation loss models for analysis. Note that, the results under UMi scenarios are not drawn because they are all above 10 Gbps or around it. As can be observed, IRR glass that is widely used in energy-saving buildings can significantly lower the PDLT and shorten communication distance. Moreover, if the human body blockage model is taken into consideration, the SNR will decrease by 30 to 40 dB and thus PDLT will be significantly lowered. As explained and illustrated in the previous sections, the DPA-MIMO architecture can mitigate the human body blockage issue by enabling the BF modules not blocked in working modes.

\begin{figure*}[!h]
\centering
\includegraphics[scale = 0.8]{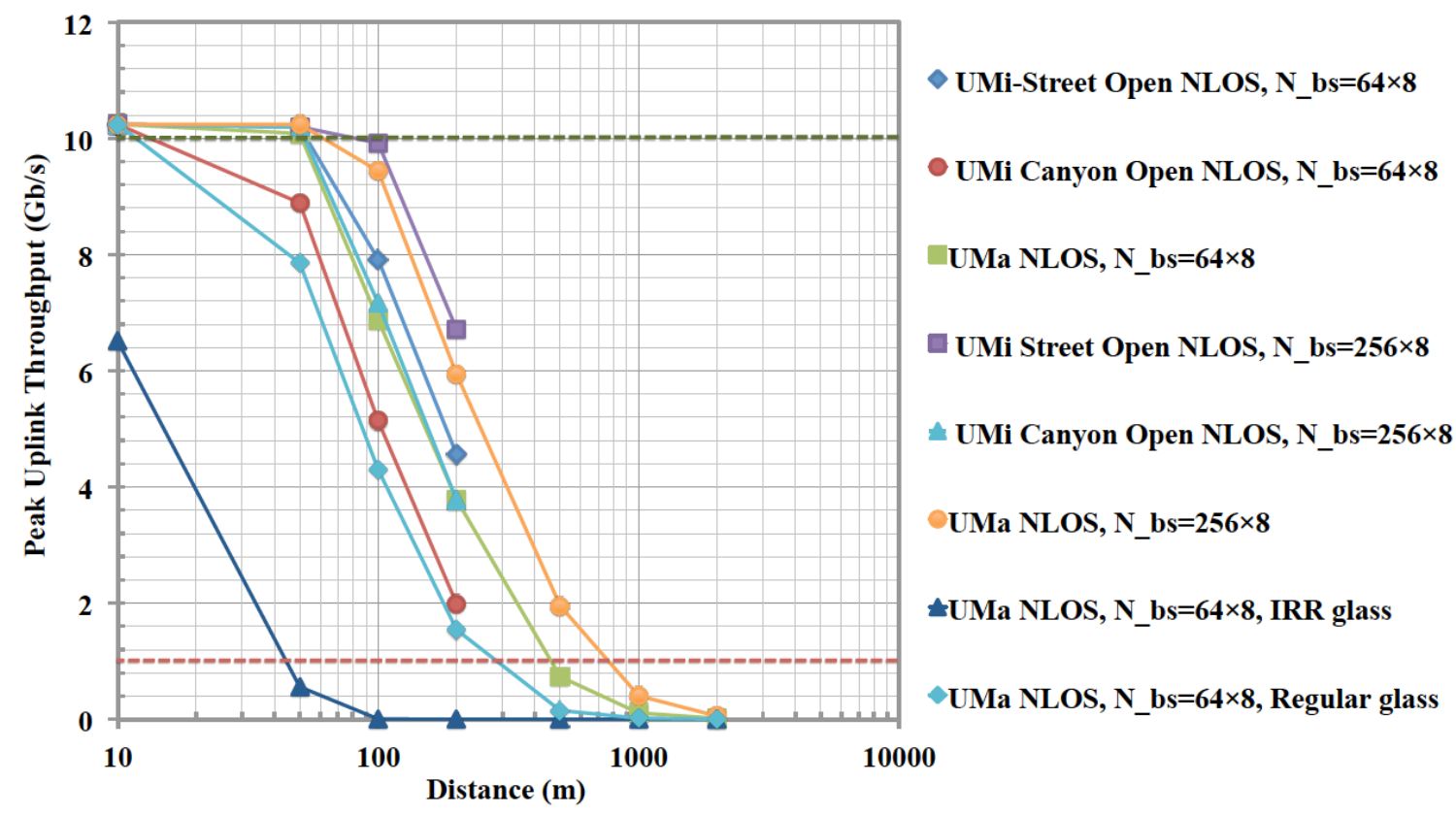}
\caption{28 GHz, $8\times8$ MIMO, BW=200 MHz, peak uplink throughput versus various deployment scenarios.}\label{fig:PULTFig}
\end{figure*}

\begin{figure*}[!h]
\centering
\includegraphics[scale =0.75]{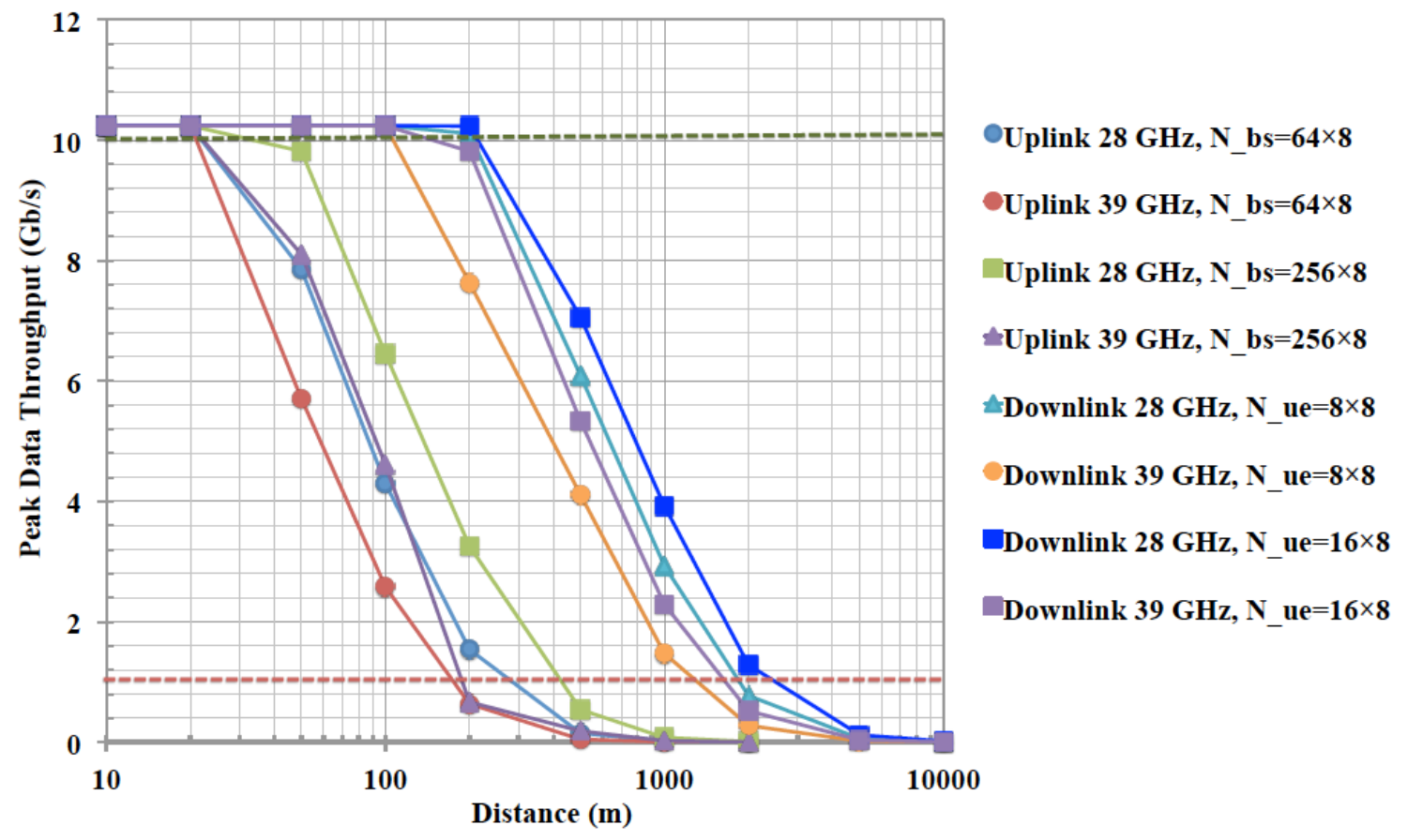}
\caption{Regular glass, UMa NLOS, $8\times8$ MIMO, BW=200 MHz, peak data throughput versus distance for various number of antenna elements on BS and UE ends.}\label{fig:GLASSFig}
\end{figure*}

\begin{figure*}[!h]
\centering
\includegraphics[scale =0.75]{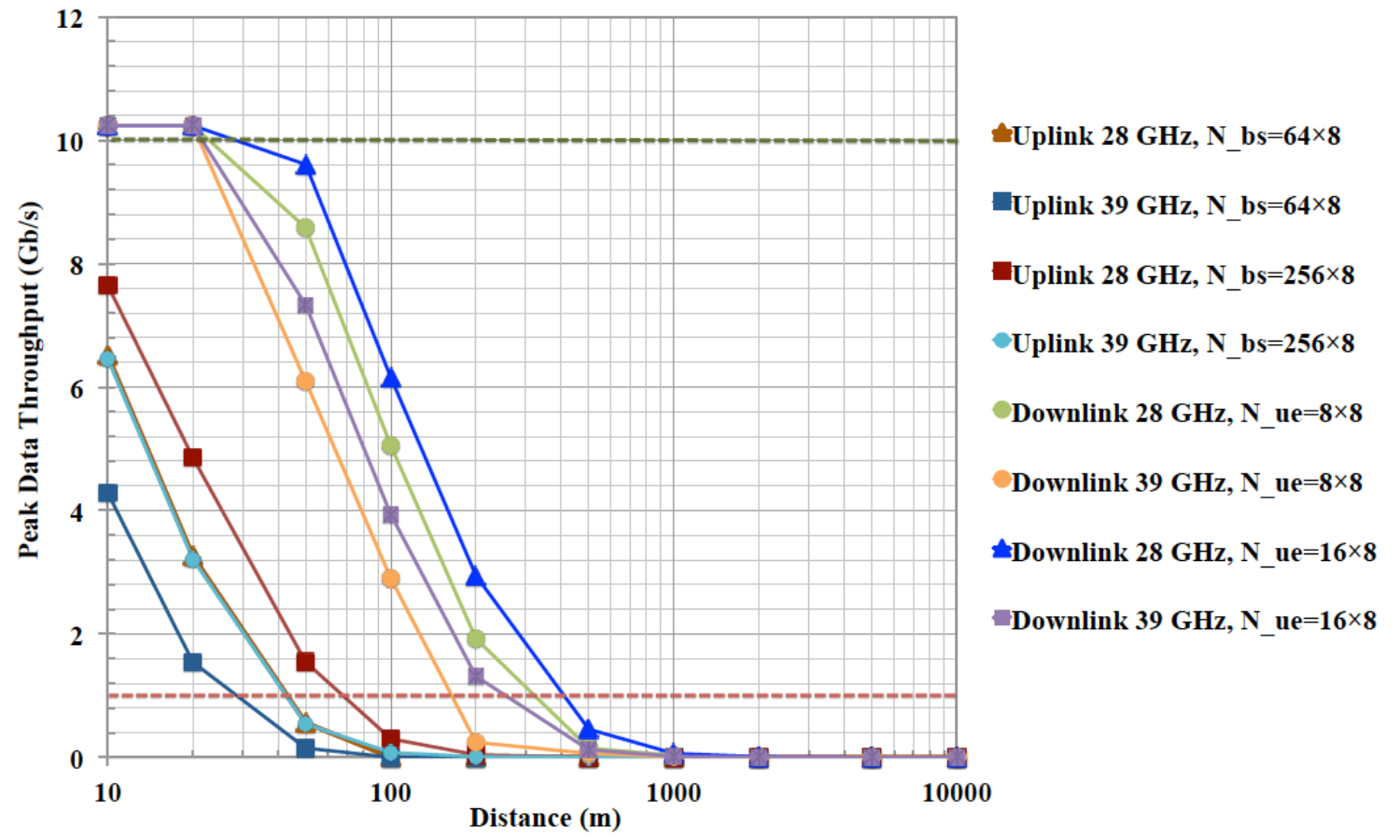}
\caption{IRR glass, UMa NLOS, $8\times8$ MIMO, BW=200 MHz, peak data throughput versus distance for various number of antenna elements on BS and UE ends.}\label{fig:IRRGLASSFig}
\end{figure*}

\subsection{Uplink Budget and Data Throughput Analysis}
For the uplink, the link budget calculation is given in Table~\ref{tab:PULT}. The maximum EIRP is regulated to be 43 dBm for mobile stations (MSs) by the FCC \cite{FCC}. Such level of transmission power is translated to a substantial challenge for long distance transmission at 5G mmWave frequency bands. Therefore there is necessity for the BS to enable the use of large antenna arrays at the receiver end to compensate for the propagation loss. Since the UE can operate with $N_\text{BF}\times{N_\text{BF}}$ MIMO to receive $N_\text{BF}$ streams simultaneously, the BS needs to enable $N_\text{BF}$ arrays of antenna elements as well, where each array of antenna elements is referred to as a base station unit. The number of antenna elements in each base station unit, $N_\text{array}$, is determined by the link budget calculation and constrained by the hardware resource and implementation feasibility on the BS end. Thus, the total number of antenna elements on the base station, $N_\text{bs}$ is equal to $N_\text{BF}\times{N_\text{array}}$. Assuming $N_\text{BF}$ is 8, two sets of data are given in Table~\ref{tab:PULT}, based on 64 and 256 antenna elements in one antenna array of the BS respectively.  

Fig.~\ref{fig:BSFig} illustrates the communication between the DPA-MIMO architecture based UE and the BS in $8\times8$ MIMO to deliver 8 streams simultaneously. When $N_\text{array}$ is 256 and the carrier frequency is 28 GHz, the dimension of the 256-antenna elements array is approximately $100\times100$ mm. In the given example, there are totally 8 such BS units. It is feasible to embed 8 or even more groups of such mmWave antenna arrays for practical hardware design of microcells and macrocells. As a matter of fact, the antenna array dimension can be further expanded to $200\times200$ mm so as to embed a 1024-antenna elements array, thereby increasing the SNR by an additional 6 dB. 

Furthermore, Fig.~\ref{fig:PULTFig} plots the peak uplink throughput (PULT) versus distance for various deployment scenarios and different numbers of antenna units on the BS end. It shows that, a large number of receiver antenna elements in the antenna array need to be enabled at the BS to compensate for the large propagation loss. Again, the IRR glass induced attenuation largely degrades the uplink performance, and therefore more antenna elements should be used at the base station to enable stronger beamforming gain. Also, at the mobile station, a maximum output of 43 dBm EIRP will limit the maximum output power of one single PA in the phased array, and the relation can be expressed in the equation below: 
 \begin{equation}\label{eq:EIRP}
\begin{aligned}
\ \text{EIRP}_{\text{UE,max}}=P_\text{PA,out}+20\text{log}_\text{10}(N_\text{ANT})
\end{aligned}
\end{equation}
where $P_\text{PA,out}$ is the output power of one single PA, and $N_\text{ANT}$ stands for the number of antenna elements in the BF module. This equation is valid only when each antenna element is connected to one PA. Therefore, when $N_\text{ANT}$  equals 16, the maximum $P_\text{PA,out}$  is limited to 19 dBm, and the maximum $P_\text{PA,out}$  will increase to 25 dBm if $N_\text{ANT}$ is set to 8. The specification of 19 dBm output power is less challenging and more implementable according to the current state-of-the-art mmWave PA designs \cite{Zhao:An}, \cite{Larie:A 60}.

\subsection{Analysis with Attenuation Models}
As previously mentioned, the IRR glass can cause very serious degradation to the uplink performance. More numerical results of data throughput under various deployment scenarios are plotted for both downlink and uplink modes with two types of penetration loss in Fig.~\ref{fig:GLASSFig} and Fig.~\ref{fig:IRRGLASSFig}, respectively. It is observed that the strong attenuation caused by IRR glass can be overcome at the cost of embedding more antennas at both the UE and BS ends. Moreover, the deployment of ultra-dense small cells is considered a necessary mechanism in mmWave communications to deal with strong path loss and attenuation \cite{DULAIMI:CELL}. 
    
On the other hand, the inter-bands carrier aggregation (Inter-CA) may be needed for 5G mmWave frequency bands. For example, 28 GHz, 37/39 GHz band, 64-71 GHz band, etc., are aggregated to provide even larger bandwidths. In that situation, the DPA-MIMO architecture can still be adopted for a practical multi-band, multi-mode 5G UE design. As a matter of fact, some multi-band 5G hardware components have been presented in the literature and can be used in the DPA-MIMO architecture. Two dual-band 5G mmWave antenna prototypes are demonstrated in \cite{Ali:Dual}, \cite{Zhai:Dual}, and a linear Doherty PA supporting 28 GH, 37 GHz and 39 GHz bands is designed and verified using SiGe process with high PAE demonstrated in \cite{Hu:A 28}. 

In order to further improve the efficiency and overcome the high peak-to-average power ratio (PAPR) issue, the power supply modulation techniques such as envelope tracking (ET) will continue to play a critical role in 5G UE design. One of the major challenges lies in the stringent signal bandwidth requirement of the ET modulator, because the supply generally tracks the signal envelope continuously, which can be many times the bandwidth of the I/Q signal components \cite{Zoya:PA}. With respect to the new 5G mmWave band occupying hundreds of MHz bandwidth, this challenge can be more pronounced. Nevertheless, some recently published work \cite{Kimball:ET} has shown a 70\% efficient envelope modulator for an X-band PA with 100-MHz signal bandwidth, which can be considered as a candidate ET technique for 5G applications.  

\begin{figure}[!h]
\centering
\includegraphics[scale = 0.5]{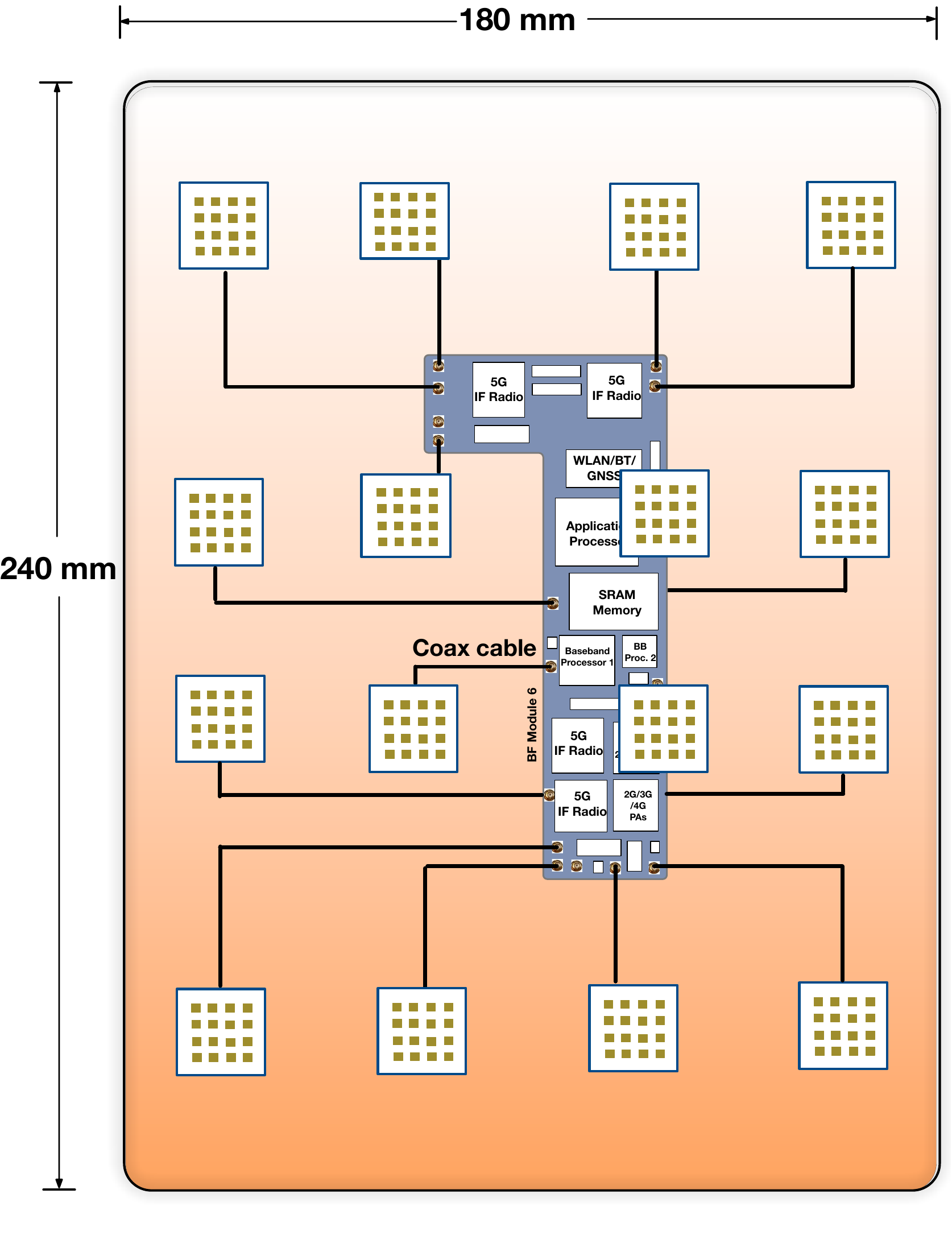}
\caption{An example of DPA-MIMO system used in tablet computer when $N_\text{ANT}$=16 and $N_\text{BF}$=16.}\label{fig:PAD17}
\end{figure}  
  
Finally, in the system-level design, the choice of $N_\text{ANT}$ and $N_\text{BF}$ can be very flexible according to the practical specification, performance target, and design constraints so that the user equipment can handle different environments and situations. For example, in Fig.~\ref{fig:5GUEFig} and Fig.~\ref{fig:5GUE16Fig}, $N_\text{ANT}$ is 8 and 16 respectively while $N_\text{BF}$ is fixed as 8, and the resulted wireless performance is summarized in Table~\ref{tab:PDLT}. On the other hand, for a specific UE design case where longer battery life and more space are available, such as in a tablet computer as shown in Fig.~\ref{fig:PAD17}, both $N_\text{ANT}$ and $N_\text{BF}$ can be increased (to 16 in Fig.~\ref{fig:PAD17}) to enable better wireless performance. 

\begin{table*}[!t]  % comparison
\scriptsize
\caption{5G state-of-the-art works performance summary and comparison } \label{tab:SoA}
\newcommand{\tabincell}[2]{\begin{tabular}{@{}#1@{}}#2\end{tabular}}
\centering
 \begin{threeparttable}
 
\newcolumntype{C}{>{\centering\arraybackslash}p{2em}} 
  \begin{tabular}{|c|c|c|c|c|c|c|c|c|c|c|c|c|c|c|c|c|}\hline 
             
        \tabincell{c}{\textbf{Item}} & \multicolumn{2}{c|}{\tabincell{c}{\textbf{Announce/}\\\textbf{Release}\\ \textbf{Date}}} & \multicolumn{2}{c|}{\tabincell{c}{\textbf{5G}\\ \textbf{Equipment} \\ \textbf{Type}}} & \multicolumn{2}{c|}{\tabincell{c}{\textbf{Hardware}\\ \textbf{Implementation}\\ \textbf{Solution and}\\ \textbf{Method}}} & \multicolumn{2}{c|}{\tabincell{c}{ \textbf{Data}\\\textbf{Resource}}} & \multicolumn{2}{c|}{\tabincell{c}{ \textbf{Air Interface}\\\textbf{Technology}}} & \multicolumn{2}{c|}{\tabincell{c}{\textbf{Carrier}\\\textbf{Frequency} \\\textbf{(GHz)}}} & \multicolumn{2}{c|}{\tabincell{c}{\textbf{Bandwidth}\\\textbf{(MHz) and} \\ \textbf{Modulation} }} & \multicolumn{2}{c|}{\tabincell{c}{\textbf{DL Throughput}\\ \textbf{and Distance}}}  \\  \hline
        
        \tabincell{c}{Samsung\\ \cite{SAMSUNG:5G_NEWS}, \cite{SAMSUNG:5G}} & \multicolumn{2}{c|}{\tabincell{c}{Oct. 2014}} & \multicolumn{2}{c|}{\tabincell{c}{BS, \\UE}} & \multicolumn{2}{c|}{\tabincell{c}{Modem board,\\analog FE board}} & \multicolumn{2}{c|}{\tabincell{c}{Outdoor\\test}} & \multicolumn{2}{c|}{\tabincell{c}{BF,\\ 2$\times$2 MIMO}} & \multicolumn{2}{c|}{\tabincell{c}{28}} & \multicolumn{2}{c|}{\tabincell{c}{800,\\64 QAM}} & \multicolumn{2}{c|}{\tabincell{c}{3.77 Gbps@20m\\ 7.5 Gbps@20m(2MSs)}}      \\  \hline
        
        \tabincell{c}{NTT-DOCOMO \\ \cite{NTT:5G}} & \multicolumn{2}{c|}{\tabincell{c}{Sep. 2016}} & \multicolumn{2}{c|}{\tabincell{c}{BS,\\UE}} & \multicolumn{2}{c|}{\tabincell{c}{BS:49-element\\array,\\UE:4-element\\array}} & \multicolumn{2}{c|}{\tabincell{c}{Outdoor\\test}} & \multicolumn{2}{c|}{\tabincell{c}{BF,\\ 2$\times$2 MIMO}} & \multicolumn{2}{c|}{\tabincell{c}{27.92}} & \multicolumn{2}{c|}{\tabincell{c}{800,\\64 QAM}} & \multicolumn{2}{c|}{\tabincell{c}{3.77 Gbps@120m}}      \\  \hline
        
        \tabincell{c}{Qualcomm\\ \cite{Qualcomm:5G_1} -\cite{Qualcomm:5G_NEWS} } & \multicolumn{2}{c|}{\tabincell{c}{Oct. 2016}} & \multicolumn{2}{c|}{\tabincell{c}{UE}} & \multicolumn{2}{c|}{\tabincell{c}{Modem SoC,\\mmWave RFIC}} & \multicolumn{2}{c|}{\tabincell{c}{IC test}} & \multicolumn{2}{c|}{\tabincell{c}{BF,\\MIMO}} & \multicolumn{2}{c|}{\tabincell{c}{28}} & \multicolumn{2}{c|}{\tabincell{c}{800,\\modulation\\(N.A.)}} & \multicolumn{2}{c|}{\tabincell{c}{5 Gbps,\\distance (N.A.)}}      \\  \hline
        
        \tabincell{c}{Intel\\ \cite{INTEL:5G}, \cite{INTEL:5G_CES} } & \multicolumn{2}{c|}{\tabincell{c}{Jan. 2017}} & \multicolumn{2}{c|}{\tabincell{c}{UE}} & \multicolumn{2}{c|}{\tabincell{c}{Modem SoC,\\mmWave RFIC}} & \multicolumn{2}{c|}{\tabincell{c}{IC test}} & \multicolumn{2}{c|}{\tabincell{c}{BF,\\MIMO}} & \multicolumn{2}{c|}{\tabincell{c}{28,\\sub-6}} & \multicolumn{2}{c|}{\tabincell{c}{800,\\modulation\\(N.A.)}} & \multicolumn{2}{c|}{\tabincell{c}{5 Gbps,\\distance (N.A.)}}      \\  \hline

         \tabincell{c}{IBM\\ \cite{IBM:5G} } & \multicolumn{2}{c|}{\tabincell{c}{Feb. 2017}} & \multicolumn{2}{c|}{\tabincell{c}{BS}} & \multicolumn{2}{c|}{\tabincell{c}{Chipsets,\\antennas in package,\\single module}} & \multicolumn{2}{c|}{\tabincell{c}{IC test,\\antenna\\chamber test}} & \multicolumn{2}{c|}{\tabincell{c}{BF}} & \multicolumn{2}{c|}{\tabincell{c}{28}} & \multicolumn{2}{c|}{\tabincell{c}{Bandwidth\\(N.A.),\\256 QAM}} & \multicolumn{2}{c|}{\tabincell{c}{3.5 Gbps@1.84 m}}      \\  \hline

           \tabincell{c}{UCSD\\ \cite{UCSD:5G} } & \multicolumn{2}{c|}{\tabincell{c}{Feb. 2017}} & \multicolumn{2}{c|}{\tabincell{c}{N.A.}} & \multicolumn{2}{c|}{\tabincell{c}{32-element \\BF module,\\ single module\\on PCB}} & \multicolumn{2}{c|}{\tabincell{c}{IC test}} & \multicolumn{2}{c|}{\tabincell{c}{BF}} & \multicolumn{2}{c|}{\tabincell{c}{28}} & \multicolumn{2}{c|}{\tabincell{c}{Bandwidth\\(N.A.),\\64 QAM}} & \multicolumn{2}{c|}{\tabincell{c}{12 Gbps,\\distance (N.A.)}     } \\  \hline
           
            \tabincell{c}{Huawei,\\NTT-DOCOMO \\ \cite{HUAWEI:5G}} & \multicolumn{2}{c|}{\tabincell{c}{Nov. 2016}} & \multicolumn{2}{c|}{\tabincell{c}{BS\\(macrocell),\\UE}} & \multicolumn{2}{c|}{\tabincell{c}{N.A.}} & \multicolumn{2}{c|}{\tabincell{c}{Field test}} & \multicolumn{2}{c|}{\tabincell{c}{Massive-\\MIMO}} & \multicolumn{2}{c|}{\tabincell{c}{4.5}} & \multicolumn{2}{c|}{\tabincell{c}{200,\\modulation\\(N.A.)}} & \multicolumn{2}{c|}{\tabincell{c}{11.29 Gbps,\\distance (N.A.)}} \\  \hline      
            \tabincell{c}{Ericsson,\\T-Mobile \\ \cite{ERICSSON:5G}} & \multicolumn{2}{c|}{\tabincell{c}{Sep. 2016}} & \multicolumn{2}{c|}{\tabincell{c}{BS,\\UE}} & \multicolumn{2}{c|}{\tabincell{c}{N.A.}} & \multicolumn{2}{c|}{\tabincell{c}{Lab test,\\field trials}} & \multicolumn{2}{c|}{\tabincell{c}{BF,\\MU-MIMO}} & \multicolumn{2}{c|}{\tabincell{c}{28}} & \multicolumn{2}{c|}{\tabincell{c}{N.A.}} & \multicolumn{2}{c|}{\tabincell{c}{12 Gbps,\\distance (N.A.)}} \\  \hline    
  
            \tabincell{c}{Nokia,\\SK Telecom\\ \cite{NOKIA_SKT:5G_NEWS}, \cite{NOKIA:5G}} & \multicolumn{2}{c|}{\tabincell{c}{Oct. 2015}} & \multicolumn{2}{c|}{\tabincell{c}{BS,\\UE}} & \multicolumn{2}{c|}{\tabincell{c}{N.A.}} & \multicolumn{2}{c|}{\tabincell{c}{PoC test}} & \multicolumn{2}{c|}{\tabincell{c}{8$\times$8 MIMO}} & \multicolumn{2}{c|}{\tabincell{c}{15}} & \multicolumn{2}{c|}{\tabincell{c}{4$\times$100,\\256 QAM}} & \multicolumn{2}{c|}{\tabincell{c}{19.1 Gbps,\\distance (N.A.)}} \\  \hline    
      
             \tabincell{c}{NI,\\Verizon\\ \cite{NI:5G}} & \multicolumn{2}{c|}{\tabincell{c}{Mar. 2017}} & \multicolumn{2}{c|}{\tabincell{c}{BS,\\UE}} & \multicolumn{2}{c|}{\tabincell{c}{NI USRP}} & \multicolumn{2}{c|}{\tabincell{c}{Demo,\\prototype test}} & \multicolumn{2}{c|}{\tabincell{c}{2$\times$2 MU-MIMO}} & \multicolumn{2}{c|}{\tabincell{c}{28}} & \multicolumn{2}{c|}{\tabincell{c}{8$\times$100,\\modulation\\(N.A.)}} & \multicolumn{2}{c|}{\tabincell{c}{5 Gbps,\\20 Gbps (scalable), \\distance (N.A.)}} \\  \hline 
             
             \tabincell{c}{\textbf{This work}} & \multicolumn{2}{c|}{\tabincell{c}{\textbf{Apr. 2017}}} & \multicolumn{2}{c|}{\tabincell{c}{\textbf{UE}}} & \multicolumn{2}{c|}{\tabincell{c}{\textbf{Multiple BF}\\\textbf{modules arranged}\\\textbf{on UE}}} & \multicolumn{2}{c|}{\tabincell{c}{\textbf{Numerical}}} & \multicolumn{2}{c|}{\tabincell{c}{\textbf{BF,}\\\textbf{8$\times$8 DPA-MIMO}}} & \multicolumn{2}{c|}{\tabincell{c}{\textbf{28}}} & \multicolumn{2}{c|}{\tabincell{c}{\textbf{800,}\\\textbf{256 QAM}}} & \multicolumn{2}{c|}{\tabincell{c}{\textbf{40.96 Gbps@100m}\\\textbf{26.49 Gbps@1000m}}} \\  \hline 
  
    \end{tabular}   
    \end{threeparttable}
\end{table*}
\subsection{Performance Comparison with State-of-the-art}
Table~\ref{tab:SoA} summarizes and compares the performance of 5G state-of-the-art works from both industry and academics in recent years. We can observe that most of the previous works have used conventional beamforming, MIMO, Massive-MIMO, and MU-MIMO as the air interface technology. According to the numerical analysis, the novel DPA-MIMO architecture not only provides the highest downlink throughput at long distances, but also is able to address the human body blockage issue which is not mentioned or dealt with in the previous 5G state-of-the-art research works. 

\section{CONCLUSION}
In this paper, a system architecture and method for next generation wireless user equipment, or 5G cellular user equipment design has been provided. By analyzing the challenges of contemporary wireless UE designs and emerging 5G mmWave techniques, a novel DPA-MIMO architecture and design method has been presented to overcome the limitations of conventional MIMO structures. This work has provided a solution to the technical constraints and challenges of the mobile handset design, such as human blockage, high path loss, self-heating issues, which are more pronounced for future 5G cellular user equipment but cannot be solved using the existing system architectures and methods. Moreover, through numerical analysis, the proposed architecture has been shown to increase the wireless link budget and enhance the data throughput under different use cases and various BS deployment scenarios with highly flexible reconfigurability. Furthermore, the DPA-MIMO based wireless UE can be implemented using the state-of-the-art technologies of circuits, antennas and systems. As a result, this architecture can facilitate a peak throughput of more than 10 Gb/s while maintaining a slim form factor of mobile terminal devices. 

\section{ACKNOWLEDGEMENT}

The authors would like to acknowledge Natural Sciences and Engineering Research Council of Canada, and National Natural Science Foundation of China for support of this project, Dr. Song Hu from Georgia Institute of Technology and Dr. Adrian Tang from NASA JPL for valuable discussions.

\begin{IEEEbiography}
{Yiming Huo}(S'08) received B.Eng degree in information engineering from Southeast University, China, in 2006, and MSc. degree in System-on-Chip (SoC) from Lund University, Sweden, in 2010. He is currently pursuing the Ph.D. in electrical engineering at University of Victoria, Canada. From 2009 to 2010, he stayed in Ericsson and ST-Ericsson where he accomplished his master thesis on multi-mode, wideband CMOS VCOs for cellular systems. Since his Ph.D. study, Mr. Huo has been working on the a wide range of cross-disciplinary topics of next generation wireless communication systems design, and his current research interests are in the physical layer design of 5G cellular user equipment, and millimeter-wave circuits and systems.

Between 2010 and 2011, Mr. Huo worked in Chinese Academy of Sciences (CAS) as research associate. From 2011 to 2012, he worked in STMicroelectronics as RF engineer for developing digital video broadcasting (DVB) systems. From 2015 to 2016, he spent a year conducting RF hardware and cellular design for various projects in Apple Inc., Cupertino, California, as a Ph.D intern.  

Mr. Huo is the recipient of the Best Student Paper Award of the 2016 \emph{IEEE ICUWB}, the Excellent Student Paper Award of the 2014 \emph{IEEE ICSICT}, and the Bronze Leaf certificate of the 2010 \emph{IEEE PrimeAsia}. He also received University of Victoria Fellowship from 2012 to 2013, and ISSCC-STGA from IEEE Solid-State Circuits Society (SSCS) in 2017.     

Mr. Huo is a member of IEEE, SSCS, CAS, MTT-S, and ComSoc. He has been serving as the program committee of IEEE ICUWB 2017, technical reviewer for several IEEE conferences and journals including \textsc{IEEE Transactions on Vehicular Technology (TVT)}, and \textsc{China Communications}.    
\end{IEEEbiography}

\begin{IEEEbiography}
{Xiaodai Dong}(S'97-M'00-SM'09) received the B.Sc. degree in information and control engineering from Xi'an Jiaotong University, Xi'an, China, the M.Sc. degree in electrical engineering from National University of Singapore, Singapore, and the Ph.D. degree in electrical and computer engineering from Queen's University, Kingston, ON, Canada, in 1992, 1995, and 2000, respectively. Since January 2005, she has been with the University of Victoria, Victoria, BC, Canada, where she is currently a Professor and a Canada Research Chair (Tier II) with the Department of Electrical and Computer Engineering. From 2002 to 2004, she was an Assistant Professor with the Department of Electrical and Computer Engineering, University of Alberta, Edmonton, AB, Canada. From 1999 to 2002, she was with Nortel Networks, Ottawa, ON, Canada, and worked on the base transceiver design of the third-generation (3G) mobile communication systems. Her research interests include mobile communications, radio propagation, ultra-wideband radio, machine to machine communications, wireless security, ehealth, smart grid, nano-communications and signal processing for communication applications. She is an Editor for the \textsc{IEEE Transactions on Vehicular Technology}.
\end{IEEEbiography}

\begin{IEEEbiography}
{Wei Xu}(S'07-M'09-SM'15) received his B.Sc. degree in Electrical Engineering in 2003 and his M.S. and Ph.D. degrees in Communication and Information Engineering in 2006 and 2009, respectively, all from Southeast University, Nanjing, China. He is currently a Full Professor at the National Mobile Communications Research Lab (NCRL), Southeast University. Between 2009 and 2010, he was a post-doctoral research fellow with the Department of Electrical and Computer Engineering, University of Victoria, Canada.

Dr. Xu is an Editor of the \textsc{IEEE Communications Letters}. He has been involved in technical program committees for many international conferences including \emph{IEEE Globecom}, \emph{IEEE ICC}, \emph{IEEE WCNC}, etc. He has published over 100 refereed journal and conference papers in addition to 6 granted patents. He received the best paper awards of \emph{IEEE MAPE} in 2013, \emph{IEEE/CIC ICCC} in 2014, and \emph{IEEE Globecom} in 2014. He was elected core team member of the Jiangsu Innovation Team in 2012. He is the co-recipient of the First Prize Award of Jiangsu Science and Technology Award in 2014. His research interests include cooperative communications, information theory and signal processing for wireless communications.
\end{IEEEbiography}

\end{document}